\documentclass{article}
\usepackage[square,numbers]{natbib}
\usepackage[preprint]{neurips_2020}

\usepackage{hyperref}
\usepackage{graphicx,multicol}
\usepackage{placeins}
\usepackage{multirow}
\usepackage{nicefrac}
\usepackage{amsmath}

\AtBeginDocument{%
  \providecommand\BibTeX{{%
    \normalfont B\kern-0.5em{\scshape i\kern-0.25em b}\kern-0.8em\TeX}}}

\begin{document}

\title{Tuning Word2vec for Large Scale Recommendation Systems}


\author{%
  Benjamin P. Chamberlain \\
   Twitter \\
   United Kingdom\\
  \texttt{bchamberlain@twitter.com} \\
   \And
  Emanuele Rossi \\
   Twitter \\
   United Kingdom\\   
   \And
   Dan Shiebler \\
   Twitter \\
   \AND
   Suvash Sedhain \\
   Twitter \\
   \And
   Michael M. Bronstein \\
   Twitter / Imperial College London\\
   United Kingdom\\   
}

\maketitle




\begin{abstract}
Word2vec is a powerful machine learning tool that emerged from Natural Language Processing (NLP) and is now applied in multiple domains, including recommender systems, forecasting, and network analysis. As Word2vec is often used off the shelf, we address the question of whether the default hyperparameters are suitable for recommender systems. The answer is emphatically no. 
In this paper, we first elucidate the importance of hyperparameter optimization and show that unconstrained optimization yields an average 221\% improvement in hit rate over the default parameters. However, unconstrained optimization leads to hyperparameter settings that are very expensive and not feasible for large scale recommendation tasks. To this end, we demonstrate 138\% average improvement in hit rate with a runtime budget-constrained hyperparameter optimization. Furthermore, to make hyperparameter optimization applicable for large scale recommendation problems where the target dataset is too large to search over, we investigate generalizing hyperparameters settings from samples. We show that applying constrained hyperparameter optimization using only a 10\% sample of the data still yields a 91\% average improvement in hit rate over the default parameters when applied to the full datasets. 
Finally, we apply hyperparameters learned using our method of constrained optimization on a sample to the Who To Follow recommendation service at Twitter and are able to increase follow rates by 15\%.

\end{abstract}







\section{Introduction}

Word2vec is a popular model for learning word representations that has since found a wide range of additional uses~\cite{Barkan2016, Wang2017a, Grbovic2018, Zhao2017b, Caselles-Dupre2018, Vasile2016a, Grbovic2015, Chamberlain2017, Perozzi2014}. Owing to its robustness, simplicity, and efficiency, Word2vec is a valuable component of many recommender systems where it is used for benchmarking, candidate generation and transfer learning~\cite{Barkan2016, Wang2017a, Grbovic2018, Zhao2017b, Caselles-Dupre2018, Vasile2016a, Grbovic2015, Ozsoy2016}. 
It is common for the default parameters given in~\cite{Mikolov2013, Rehurek2010} to be taken (e.g. \cite{Barkan2016, Grbovic2015}), but we demonstrate that these are a poor choice for recommender systems. 

The most related work to ours is~\cite{Caselles-Dupre2018}, which is a valuable contribution to the field, but has three  major limitations. Firstly only small datasets of up to roughly one million tokens are used. Word2vec is designed for the big data domain and the extent to which hyperparameters learned on small datasets generalize is not addressed~\footnote{For instance the original Word2vec authors point out that the optimal number of negative samples is dataset size dependent~\cite{Mikolov2013}}. Secondly, the results are achieved through unconstrained optimization by using orders of magnitude more computational power than is required to run with the default hyperparameters (e.g. by using 22-30 \textit{times} as many epochs). For large scale systems operating under constraints this is generally not possible. Finally, there are no online experiments showing.

For real-world utility, the following questions must be addressed  1) How much improvement in performance can be expected for large scale systems with constrained computational resources? 2) How do you select hyperparameters when the target dataset is too large to optimize over directly? 3) Can these improvements generalize to production systems?
To this end we first establish benchmarks for unconstrained optimization.
In all experiments, optimizing the hyperparameters without constraints leads to models that require orders of magnitude more computational resources to run. 
We then constrain optimization by fixing the runtime to the default runtime (with fixed hardware) and discover that on average over 60\% of the unconstrained gains are still possible without increasing computational resources. 
As parallelization of Word2vec using distributed computing frameworks like Apache Spark is challenging~\cite{Ordentlich}, for large scale systems, a single training run may take days to complete and hyperparameter optimization either with parallelized random / grid search or sequential Bayesian optimization exceeds resource constraints. 
One solution is to sample the data and run hyperparameter optimization on the sample, but there are no guarantees that the settings discovered on the sample will generalize to the full dataset.
We show that results from a small sample to generalize well for larger datasets, both in the offline case when training on 10\% of the data and in online testing when the hyperparameter search is performed on a $\nicefrac{1}{2000}$ sample of the data.
Our main contributions are:
\begin{enumerate}
\item  Show substantial improvements in recommendation quality are possible through hyperparameter optimization.
\item Demonstrate that substantial improvements in recommendation quality are possible without increasing computational resources and so results are readily applicable to large scale systems.
\item Establish that good hyperparameters can be estimated from a sample by performing constrained optimization and evaluate this in online tests on the Twitter Who To Follow recommender system.
\end{enumerate}

\section{Word2vec Background}
\label{sec:skipgram}

There are two Word2vec~\citep{Mikolov2013} models, Skipgram and the Continuous Bag of Words (CBOW). They are trained self-supervised on sequences of tokens with a loss function that is based on the categorical cross-entropy of an approximation to the softmax function. There are two ways to approximate the softmax 1) negative sampling~\cite{Mnih2012} 2) hierarchical softmax~\cite{Mnih2008}. Here we use negative sampling as it is faster and achieves similar results.

\begin{figure*}
\centering
\includegraphics[width=0.75\textwidth]{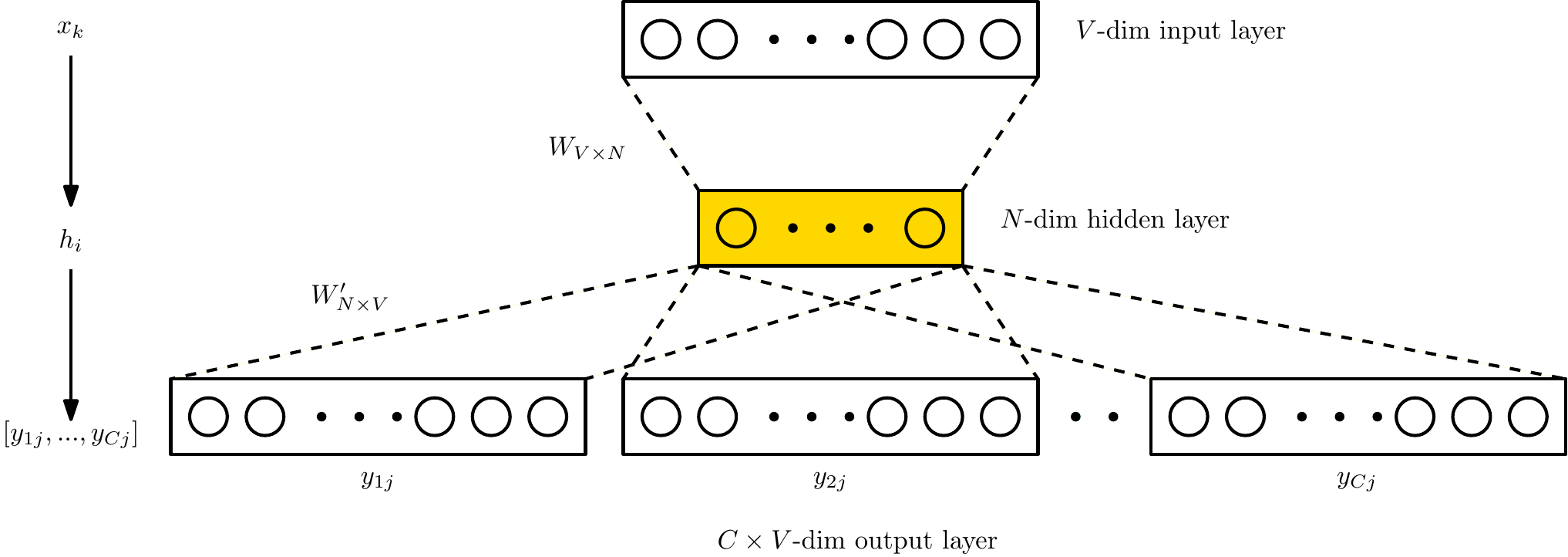}
\caption[The Word2vec architecture]{The Word2vec architecture has three layers: An input projection layer, a linear hidden layer and a softmax output.}
\label{fig:skipgram}
\end{figure*}

CBOW and Skipgram (See Figure~\ref{fig:skipgram}) are shallow neural network models. In both, two dense vector representations are learned for each entity, one as a target ($W$ in Figure~\ref{fig:skipgram}) and one as a context object ($W'$ in Figure~\ref{fig:skipgram}). Once trained, the output layer is usually discarded and the input-embedding is treated as the vector representation.
%



The five most important Word2vec hyperparameters are shown in Table~\ref{tab:hyperparams}. They are 1) The dimension of the learned vectors $d$. This determines the number of parameters of the model 2) The maximum sliding window size $L$, which determines the maximum separation in a sequence of two tokens that can interact in training 3) the negative sampling exponent $\alpha$ where $\alpha=0$ leads to uniform sampling of entities, $\alpha=1$ to unigram or popularity sampling and $\alpha=-1$ to inverse popularity sampling 4) the number of negative samples $N$ to use for each context, target pair 5) the initial learning rate (Word2vec uses linear learning rate decay). In addition, Word2vec requires a downsampling threshold and a minimum entity occurrence threshold. We set the minimum entity occurrence to one and downsampling ratio to $10^{-5}$. 

\paragraph{Downsampling}
\label{sec:downsampling}

Word2vec uses a downsampling parameter to reduce the frequency of the most frequent words. A lower value of this parameter causes more downsampling. The probability of keeping and token $x_i$ is given by 
\begin{align}
    p(x_i) = \sqrt{\frac{f}{t_i+1}}\left( \frac{t}{f_i}\right ),
    \label{eq:downsampling_prob}
\end{align}
where $t$ is the downsampling threshold ratio (typically in $[0,10^{-3}]$) times the number of tokens and $f_i$ is the number of tokens of type $i$. While Equation~\eqref{eq:downsampling_prob} is not strictly a probability, tokens are removed whenever
\begin{align}
    p(x_i) < rand(0,1),
\end{align}
where rand(0,1) generates uniform random samples in the interval [0,1] and so it only applies when $f_i \geq t$. This is a strange distribution and in the original paper~\cite{Mikolov2013} the simpler $p(x_i)=\sqrt{t/f_i}$ was used. However, this is the distribution in both the gensim and the word2vec.c implementations, presumably as it gives better performance on the original datasets. Downsampling is not simply for runtime optimization and we find that for some datasets using $1e^{-5}$ instead of default $1e^{-3}$ leads to up to $50\%$ improvements in hit rate. This is because removing frequent tokens a-priori extends the effective window size allowing tokens that would have been too distant without downsampling to interact.

While downsampling is an important parameter, we found that setting the downsampling threshold to $1e^{-5}$ works well for all datasets, with minimal improvement possible from additional optimization. 

\paragraph{Window Size}

The parameter $L$ determines the maximum sequence distance between two tokens that can interact. 
It is often overlooked that the hyperparameter $L$ is the maximum window size and at each iteration the window size $l$ is sampled uniformly from $(0,1, ... , L)$. When $l$ exceeds the length of the sequence, all available tokens are used. We experimented with removing window size sampling and discovered that our models tended to overfit with validation performance dropping after an optimal number of epochs. We hypothesize that sampling the window size is vital to ensuring training sample diversity and leave formal verification of this hypothesis for future work.  

\paragraph{Runtime Complexity}

To compare the runtime complexity of CBOW and Skipgram, we define a single iteration to be all calculations that are performed for a given context window and target pair, (shown in Figure~\ref{fig:skipgram} for window size three).  
Skipgram and CBOW are very similar in structure, but while Skipgram performs three independent updates in Figure~\ref{fig:skipgram} for every $(x,y_i)$ pair, CBOW performs one by taking the mean of the $y_i$ in the hidden layers. 
An iteration of CBOW has a runtime complexity of $O(d(L+N))$. By contrast, while Skipgram is $O(LdN)$. Consequently, CBOW runs approximately three times faster than Skipgram with the default hyperparameters and can run orders of magnitude faster for large $L$ and $N$. 

\begin{table}
\begin{center}
\scalebox{0.9}{
\begin{tabular}{ cccc } 
 \hline
 \textbf{Parameter} & \textbf{Default} & \textbf{Description} \\ 
 \hline
 $d$ & 100 &  embedding vector dimension \\ 
 $L$ & 5 & sliding window max length \\ 
 $\alpha$ & 0.75 & negative sampling exponent \\
 $N$ & 5 & number of negative samples \\ 
 $\lambda$ & 0.025 & initial learning rate \\
 \hline
\end{tabular}
}
\caption{Word2vec hyperparameters and default values}
\label{tab:hyperparams}
\end{center}
\vspace{-8mm}
\end{table}

\section{Datasets}

The experimental datasets are described in Table~\ref{tab:dataset_stats}. In all cases, a sequence corresponds to the actions of a single user sorted in time. 30Music~\cite{Turrin2015} and Deezer~\cite{Caselles-Dupre2018} are sequences of tracks in playlists from internet radio stations. Ecom~\cite{Chen2012} is transactions from a UK-based online retailer taken over one year. Kosarak~\cite{Bodon2003} is a click-stream from a Hungarian online news service. Twitter follows are sequences of user follow events occurring between February 2019 and February 2020 with the test set drawn from March 2020. Each sequence corresponds to a single user and the tokens are the IDs of users that they followed in that year. Twitter Retweet is a subset of the 2020 RecSys Challenge dataset~\cite{Belli2020}. It is generated from the full RecSys Challenge data by selecting only users that have retweeted between five and 100 times. Each sequence corresponds to a single user and the tokens are the IDs of users that they have retweeted. The Twitter retweet training set is drawn from a single week in 2020 with the test set originating from the following week.

\begin{table*}[!tbh]
\vspace{-2mm}
\begin{center}
\scalebox{0.9}{
\begin{tabular}{ ccccccccc } 
 \hline
 \textbf{Dataset} & \% & \textbf{\#entities} & \textbf{\#sequences} & \textbf{seq stats} & \textbf{\#tokens} & \textbf{ratio} & \multicolumn{2}{c}{\textbf{default performance}} \\ \hline & & & & & & & \textbf{HR@10} & \textbf{NDCG@10}  \\
 \hline
 \multirow{2}{*}{30Music} & 100 & 510531 & 100000 & 6, 10, 10.9, 20 & 1095964 & 2.1 & 3.50 $\pm 0.11$ & 0.019 $\pm$ 0.000 \\ 
    & 10 & 73349 & 10000 & 5, 9, 10.0, 19 & 99762 & 1.4 &&\\ 
    \hline
 \multirow{2}{*}{Deezer} & 100 & 338509 & 100000 & 3, 8, 12.1, 50 & 1210669 & 3.6  & 4.53 $\pm$ 0.10 & 0.026 $\pm$ 0.001 \\ 
   & 10 & 57766 & 10000 & 2, 7, 11.1, 49 & 110732 & 1.9 &&\\
  \hline
 \multirow{2}{*}{Ecom} & 100 & 3684 & 4234 & 3, 44, 96.0, 7983 & 406632 & 110.4 & 20.23 $\pm$ 0.29 & 0.137 $\pm$ 0.002 \\
   & 10 & 2933 & 423 & 2, 41, 84.4, 1639 & 35718 & 12.2 &&\\
  \hline
 \multirow{2}{*}{Kosarak}  & 100 & 22985 & 83625 & 2, 4, 9.4, 2108 & 787268 & 34.3 & 3.20 $\pm$ 0.25 & 0.020 $\pm$ 0.001\\ 
   & 10 & 8887 & 8362 & 1, 3, 8.4, 633 & 70292 & 7.9 &&\\ 
  \hline
 Twitter & 100 & 41040 & 291020 & 9, 21, 35.4, 499 & 10302354 & 251.0  & 10.47 $\pm$ 0.029 & 0.062 $\pm$ 0.001 \\
  follow & 10 & 40953 & 29102 & 9, 21, 35.5, 495 & 1032981 & 25.2 &&\\
   \hline
 Twitter  & 100 & 672528 & 395149 & 3, 5, 8.3, 98 & 3691174 & 5.5 & 4.45 $\pm$ 0.39 & 0.023 $\pm$ 0.002 \\
  retweet & 10 & 87838  & 39514 & 2, 4, 7.4, 96 & 292242 & 3.3 && \\
 \hline
\end{tabular}
}
\caption{Dataset statistics. The seq stats column gives statistics on the lengths of sequences in the format min,median,mean,max. Ratio is the number of tokens per entity. The \% column indicates whether the row refers to the full dataset or the 10\% sample. In all cases 10\% of the sequences were sampled rather than sampling user IDs.}
\label{tab:dataset_stats}
\end{center}
\end{table*}

\section{Methodology}

We follow the experimental design used in~\cite{Caselles-Dupre2018} and use the task of next event prediction to evaluate recommendation performance. The test set consists of pairs of query and target tokens with the goal being to predict the target from the query. The predictions are the 10 nearest vectors to the query token's vector ordered by cosine similarity and we evaluate HR@10 and NDCG@10.
We use the open-source software Gensim~\cite{Rehurek2010} for training. 
In addition we make several improvements to the setup in~\cite{Caselles-Dupre2018}. We discard identical query and prediction tokens instead of treating them as an automatic hit as a recommendation system would never have this behavior. Finally, instead of using a library that dynamically selects nearest neighbor algorithms depending on the dataset properties we use the same approximate nearest neighbor algorithm based on cosine similarity for all experiments.
For the two Twitter datasets, timestamps are available and so training and test sets are formed using a stratified temporal split. Full sequences are used for the training set and only the last two tokens from sequences occuring during the test period are used for the test set. The other datasets do not contain timestamps and so the training set is taken to be all, but the last token of each sequence and the test set the final two tokens, thus ensuring that all test query vectors are contained within the training set. 

Hyperparameter search combines Sobol random search~\cite{Antonov1979} and Gaussian process Bayesian optimization with the expected improvement acquisition function~\cite{Melorose2015}. All searches used Twitter's Tuner software. We initialized the Gaussian process with nine steps of Sobol random search before performing Bayesian optimization to convergence.

\section{Experimental Results}

Our experiments address four research questions. RQ1: How much recommendations improvement can be expected from unconstrained optimization of Word2vec hyperparameters? RQ2: How much improvement can be expected by optimization under the constraint that the computational resources are fixed? RQ3: How well can the optimal hyperparameters be estimated from a sample of the data? RQ4: How much improvement can be expected from hyperparameter optimized Word2vec in an online experiment to improve Who To Follow recommendations at Twitter?


\subsection{Unconstrained Optimization}
\label{sec:unconstrained}

Table~\ref{tab:opt} reports optimal settings of Word2vec hyperparameters and the answer to RQ1 is that an average of 221\% and 210\% improvement in HR@10 and NDCG@10 can be achieved by unconstrained hyperparameter estimation.  The search space was $d \in (10,500)$, $L \in (1,200)$, $\alpha \in (-1,1)$, $\lambda \in (0.001, 0.1)$ and $N \in (1, 200)$. In all cases models ran for approximately 100 epochs before convergence, defined as ten epochs with less than 0.01 improvement in HR@10~\footnote{we found that models often appear to converge and performance even decreases for several epochs before breaking through to new highs.}. With the exception of Twitter retweets (where the difference is small and inside the 95\% confidence bounds), Skipgram always outperformed CBOW. The columns $\%HR\uparrow$ and $\%NDCG\uparrow$ show the percentage improvement above the default parameter performance (for default performance values refer to Table~\ref{tab:dataset_stats}). The results show that, if possible, models should be run to convergence with parameters that imply far more computation than the defaults. 

\begin{table*}
\begin{center}
\scalebox{0.9}{
\begin{tabular}{ ccccccccccc } 
 \hline
 \textbf{Dataset} & \textbf{m} & \textbf{d} & \textbf{L} & \textbf{$\alpha$} & \textbf{$\lambda$} & \textbf{N} & \textbf{HR@10} & \textbf{NDCG@10} & \textbf{\%HR $\uparrow$} & \textbf{\%NDCG $\uparrow$} \\
 \hline
 30Music & sg & 115 & 6 & -0.34 &  0.040 & 8 & 14.77 $\pm 0.13$ & 0.085 $\pm$ 0.001 & 322 & 345 \\ 
 Deezer & sg & 50 & 3 & -.05  &  0.025& 5 & 11.11 $\pm$ 0.59 & 0.063 $\pm$ 0.003 & 145 & 144 \\ 
 Ecom & sg & 75 & 4 & 0.67  & 0.100 & 9 & 25.8 $\pm$ 0.43 & 0.167 $\pm$ 0.004 & 28 & 22 \\
 Kosarak & sg & 178 & 90 & -1 & 0.016 & 131 & 26.55 $\pm$ 1.2 & 0.150 $\pm$ 0.014 & 730 & 650 \\ 
 Twitter follow & sg & 326 & 103 & -0.27 & 0.015 & 80 & 16.95 $\pm$ 0.21 & 0.102 $\pm$ 0.094 & 62 & 65 \\
 Twitter RT  & cbow & 279 & 12 & 0.36 & 0.04 & 2 & 6.09 $\pm$ 0.18 & 0.031 $\pm$ 0.001 & 37 & 35 \\
 \hline
 \end{tabular}
 }
 \caption{Unconstrained optimisation. Performance improvement defined as the relative percentage above the default parameters. $m$ is the model. Models run to convergence, 95\% confidence intervals reported for HR and NDCG over five runs}
\label{tab:opt}
\end{center}
\end{table*}

\subsection{Constrained Optimization}
\label{sec:constrained}

Table~\ref{tab:constrained} reports optimal hyperparameters for models using fixed computational resources and the answer to RQ2 is that without increasing runtime above the runtime of the default model average improvements of 138\% and 120\% in HR@10 and NDCG@10 respectively can be achieved. In these experiments the number of epochs $n$ for each run was calculated to be the largest value such that the runtime was less than a run with the default hyperparameters. The same hardware and number of workers was used in all experiments. A tighter search space than in Section~\ref{sec:unconstrained} was used reflecting that constrained optimization requires a trade-off between epochs, window size, embedding dimension and negative samples. The search space was $d \in (10,200)$, $L \in (1,40)$, $\alpha \in (-1,1)$, $\lambda \in (0.001, 0.1)$ and $N \in (1, 40)$. As CBOW is more data efficient than Skipgram, particularly for high $L$ it is surprising that, with the exception of Twitter retweets, Skipgram always performs best when runtime is fixed.

\begin{figure*}[!tbh]
\vspace{-6mm}
\begin{multicols}{2}
    \includegraphics[width=\linewidth]{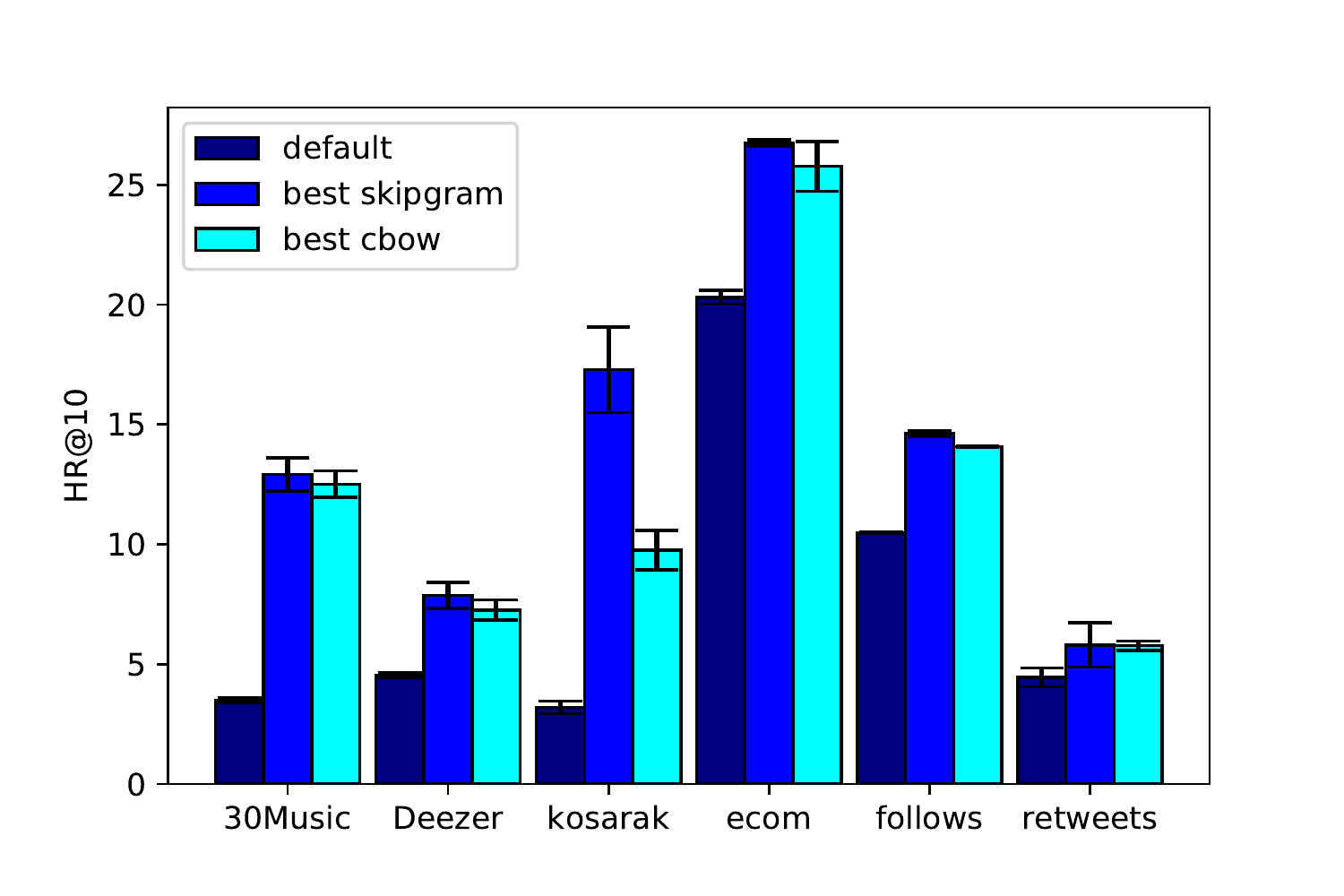} \par
    \includegraphics[width=\linewidth]{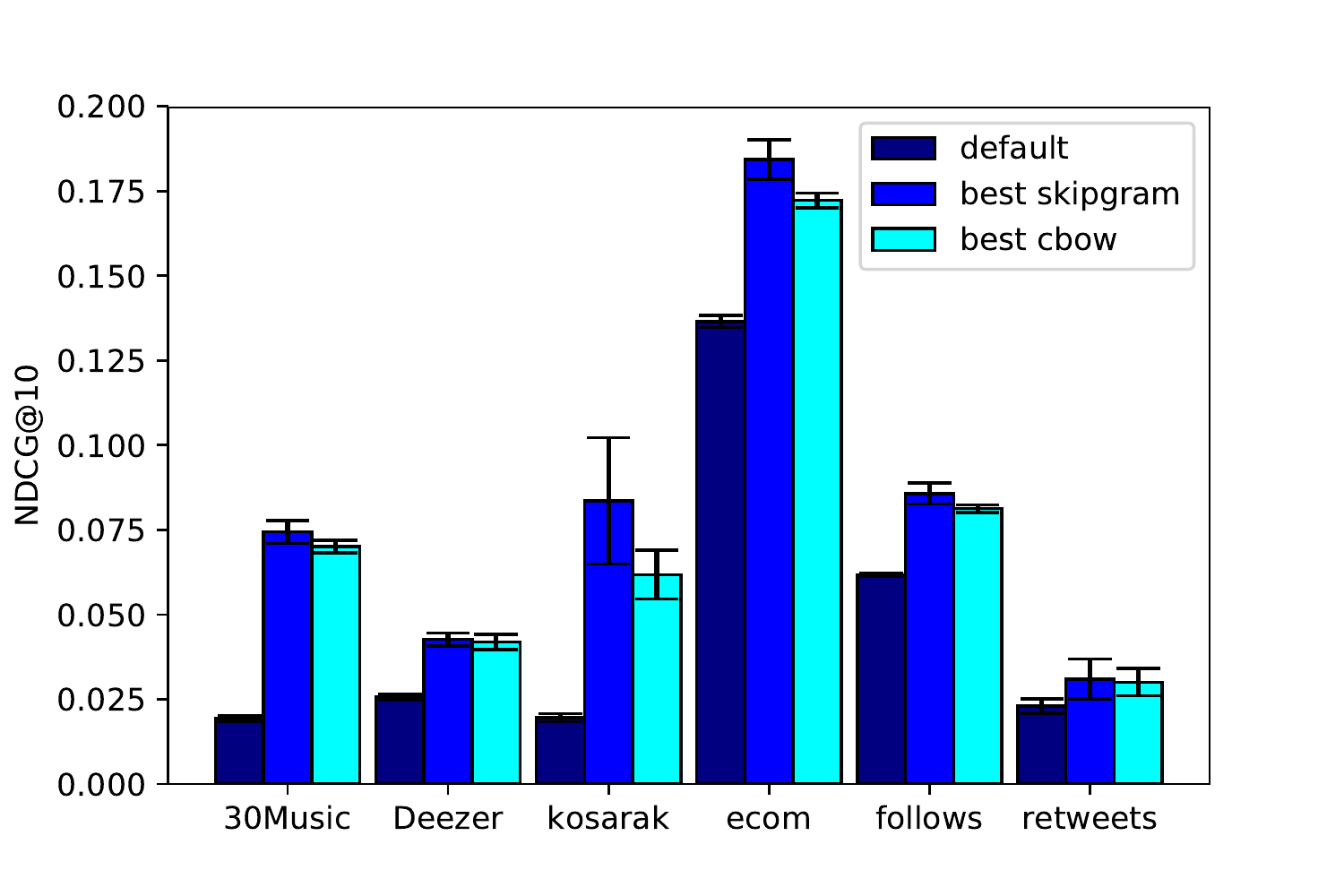} \par
\end{multicols}
\vspace{-8mm}
\caption{Comparing the performance of models with a fixed runtime budget. In all cases the optimized models were constrained to have a runtime no greater that the runtime of the model using the default parameters.}
\label{fig:constrained}
\end{figure*}

\begin{table*}[!bth]
\begin{center}
\scalebox{0.9}{
\begin{tabular}{ ccccccccccccc } 
 \hline
 \textbf{Dataset} & \textbf{m} & \textbf{d} & \textbf{L} & \textbf{$\alpha$} & \textbf{n}  & \textbf{$\lambda$} & \textbf{N} & \textbf{HR@10} & \textbf{NDCG@10} & \textbf{\%HR $\uparrow$} & \textbf{\%NDCG $\uparrow$} \\
 \hline
 \multirow{2}{*}{30Music} & sg & 36 & 7 & -0.18 & 10 & 0.092 & 6 &  12.92 $\pm 0.69$ & 0.074 $\pm$ 0.002 & 269 & 289 \\ 
  & cbow & 141 & 8 & -0.14 & 17 & 0.100 & 9 & 12.51 $\pm 0.55$ & 0.07 $\pm$ 0.002 & 257 & 289 \\ 
   \hline
 \multirow{2}{*}{Deezer} & sg & 40 & 8 & 1.0 & 13 &  0.055 & 2 & 8.35 $\pm$ 0.45 & 0.048 $\pm$ 0.003  & 84 & 85 \\ 
 & cbow & 64 & 10 & 1.0 & 14 &  0.061 & 7 & 7.77 $\pm$ 0.25 & 0.045 $\pm$ 0.003 & 72 & 73 \\ 
  \hline
 \multirow{2}{*}{Ecom} & sg & 46 & 3 & 0.39 & 7 & 0.100 & 6 & 26.73 $\pm$ 0.63 & 0.184 $\pm$ 0.006 & 32 & 34 \\
  & cbow & 87 & 8 & 0.04 & 11 & 0.027 & 8 & 24.20 $\pm$ 0.66 & 0.174 $\pm$ 0.003 & 20 & 27 \\
   \hline
 \multirow{2}{*}{Kosarak} & sg & 46 & 8 & -0.44 & 3 & 0.050 & 7 & 15.83 $\pm$ 1.71 & 0.074 $\pm$ 0.016  & 395 & 270 \\ 
   & cbow & 62 & 7 & -0.54 & 8 & 0.048 & 8 & 10.06 $\pm$ 0.09 & 0.054 $\pm$ 0.004 & 214 & 170 \\ 
    \hline
 \multirow{2}{*}{Twitter follow} & sg & 168 & 10 & 0.80 & 6 & 0.098 & 3 & 14.61 $\pm$ 0.14 & 0.086 $\pm$ 0.003 & 40 & 39 \\
  & cbow & 120 & 10 & 0.21 & 8 & 0.038 & 8 & 14.07 $\pm$ 0.25 & 0.081 $\pm$ 0.001  & 34 & 31 \\
   \hline
 \multirow{2}{*}{Twitter RT} & sg  & 25 & 31 & 0.76 & 2 & 0.030 & 14 & 4.69 $\pm$ 0.48 & 0.024 $\pm$ 0.003  & 5 & 4\\
  & cbow  & 83 & 16 & -0.21 & 7 & 0.049 & 18 & 4.74 $\pm$ 0.51 & 0.024 $\pm$ 0.001 & 7 & 4 \\
 \hline
 \end{tabular}
 }
 \caption{Constrained optimization. Optimal hyperparameters when runtime is constrained to the runtime using the default parameters for $n$ epochs, 95\% confidence intervals reported for HR and NDCG over five runs}
 \label{tab:constrained}
\end{center}
\end{table*}
Figure~\ref{fig:constrained} compares the model performance using the default hyperparameters with the best CBOW and best Skipgram models for both NCDG@10 and HR@10. With the exception of Twitter retweets, Skipgram is the best performing model and optimization significantly improves the results. This is surprising, particularly for the Twitter follows data as for fixed hyperparameters, when large windows are used, the runtime for Skipgram is much greater than CBOW.
%

\subsection{Constrained Optimization on a Sample}

\begin{figure*}[t]
\begin{multicols}{2}
    \includegraphics[width=\linewidth]{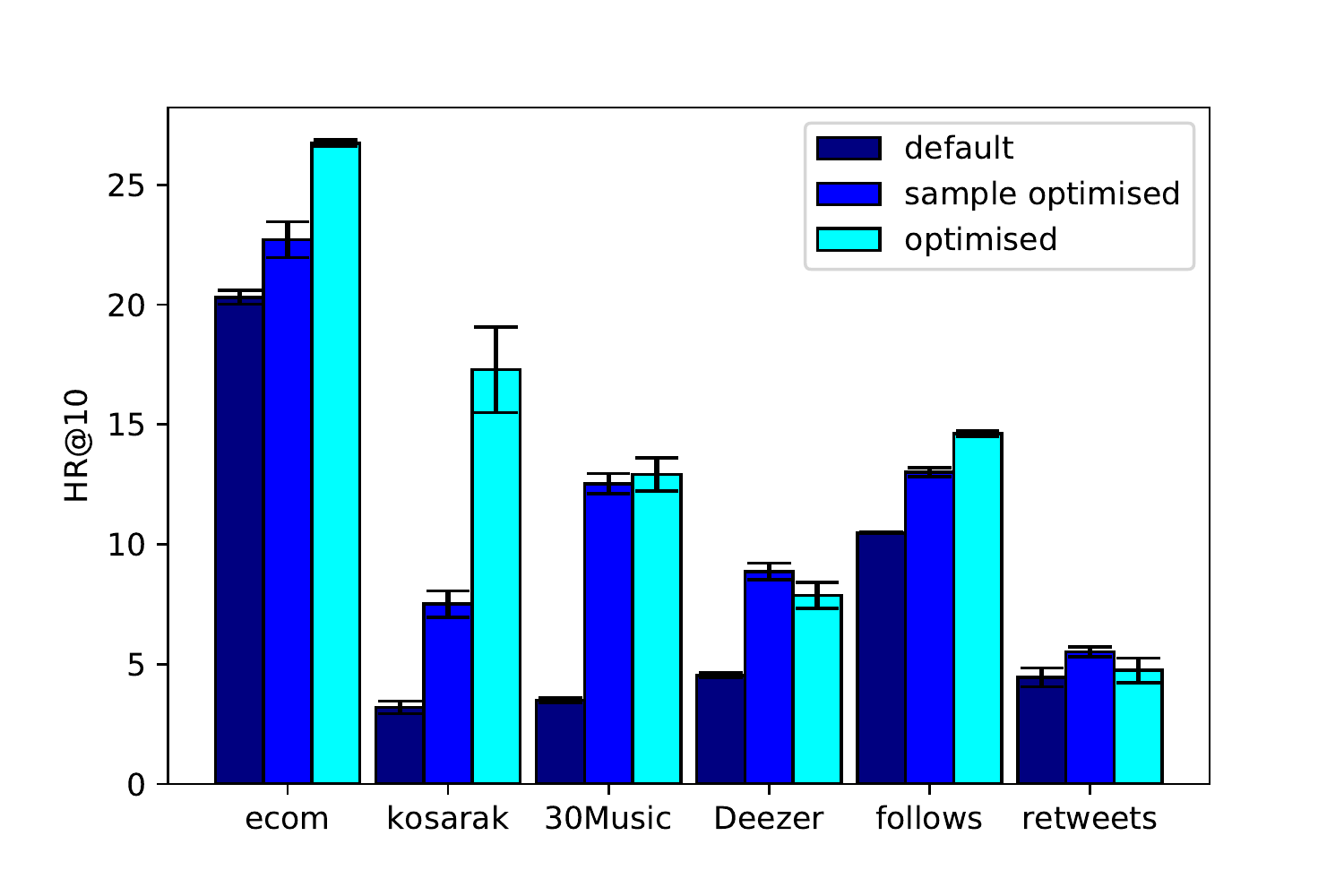} \par
    \includegraphics[width=\linewidth]{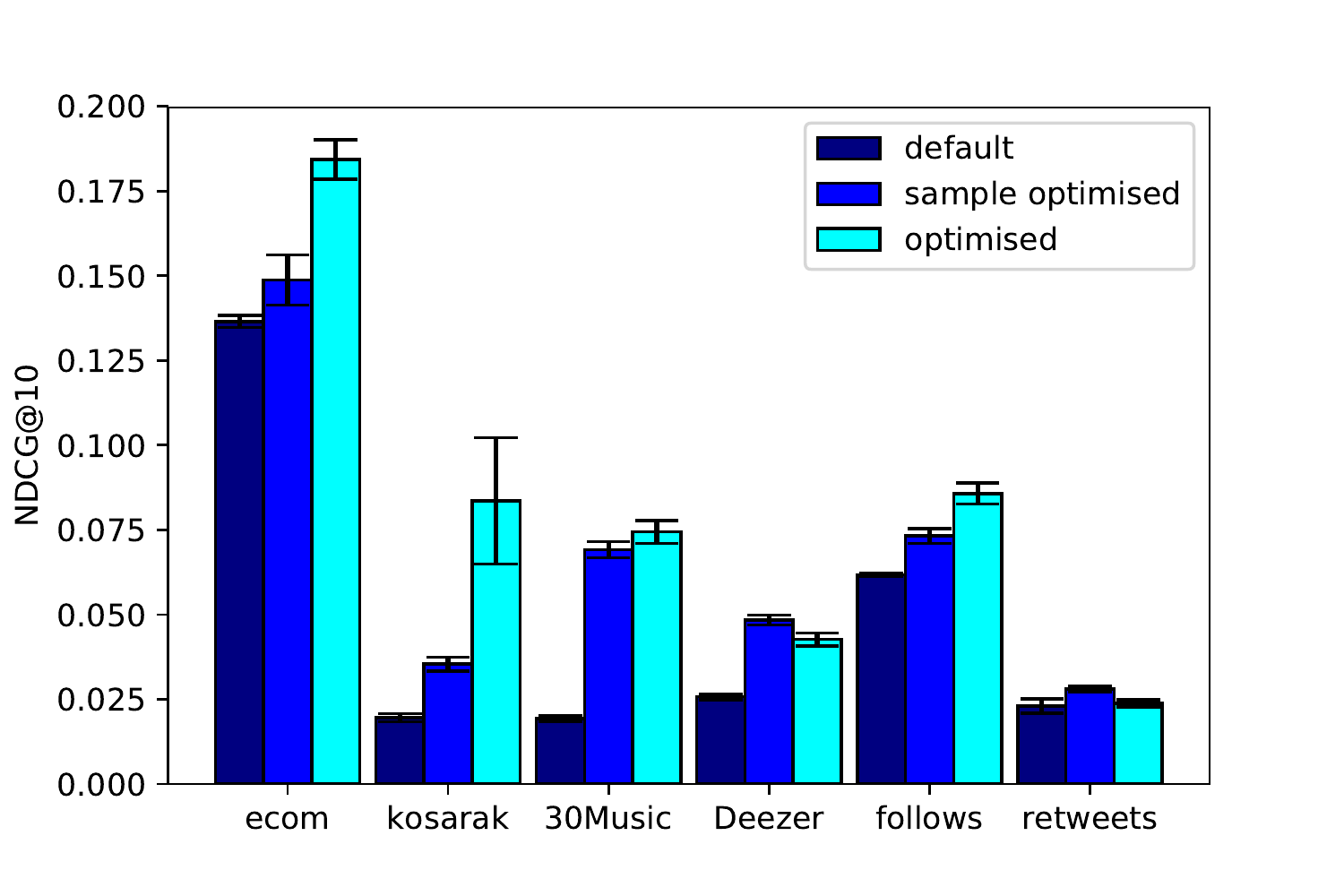} \par
\end{multicols}
\vspace{-8mm}
\caption{HR@10 with default, learned from a 10\% sample, and learned using the full dataset parameters. Optimized models constrained to have a runtime no greater that the runtime using the default parameters.}
\label{fig:dataset_size}
\vspace{-5mm}
\end{figure*}

The sampled datasets were generated by randomly sampling 10\% of the sequences and properties of the sampled datasets are given in Table~\ref{tab:dataset_stats}. The same search space and method for constraining the runtime was used as in Section~\ref{sec:constrained}.
Table~\ref{tab:sample} reports optimal hyperparameters discovered using a 10\% sample of the original datasets, but with HR@10 and NDCG@10 evaluated on the full datasets. Figure~\ref{fig:dataset_size} compares the performance on the full datasets of the hyperparameters estimated using a sample and the full dataset. The answer to RQ3 is that most of the benefit of hyperparameter optimization can be realized by doing optimization on a sample with an average of $91 / 128 \%$ and $79 / 120 \%$ for HR@10 and NDCG@10 respectively. For four of the six datasets, most of the benefit from optimization can be achieved by running hyperparameter search on a sample of the data. The exceptions are the Ecom and Kosarak datasets. Table~\ref{tab:dataset_stats} shows that the distribution of sequence lengths for these datasets is much more skewed than the others and more careful sampling, or preprocessing to normalize input sequences, is likely to be required to improve sample estimation in these cases.
We found that for Kosarak, Deezer and Twitter follows the optimal CBOW was the best performing model when evaluated on the sample, but the optimal Skipgram performed far better when evaluated on the full datasets. For this reason, when hyperparameter optimization is only possible on a sample, it is important to evaluate both the best Skipgram and best CBOW models on the full dataset. 

\begin{table*}[!]
\begin{center}
\begin{tabular}{ ccccccccccc } 
 \hline
 \textbf{Dataset} & \textbf{m} & \textbf{d} & \textbf{L} & \textbf{$\alpha$} & \textbf{n}  & \textbf{$\lambda$} & \textbf{N} & \textbf{HR@10} & \textbf{NDCG@10}  \\
 \hline
 \multirow{2}{*}{30Music} & sg & 97 & 8 & 0.35 & 24 & 0.100 & 1 &  12.53 $\pm 0.42$ & 0.069 $\pm$ 0.002  \\ 
  & cbow & 72 & 8 & -1.00 & 25 & 0.100 & 4 & 9.53 $\pm 0.10$ & 0.053 $\pm$ 0.003  \\ 
  \hline
 \multirow{2}{*}{Deezer} &  sg & 64 & 8 & -0.15 & 16 &  0.067 & 2 & 8.86 $\pm$ 0.08 & 0.035 $\pm$ 0.001  \\ 
 & cbow & 143 & 7 & -0.69 & 15 &  0.100 & 10 & 5.78 $\pm$ 0.17 & 0.033 $\pm$ 0.002 \\ 
 \hline
 \multirow{2}{*}{Ecom} & sg & 26 & 10 & 1.00 & 6 & 0.077 & 2 & 22.69 $\pm$ 0.75 & 0.149 $\pm$ 0.007  \\
  & cbow & 141 & 2 & 0.24 & 15 & 0.029 & 4 & 20.43 $\pm$ 0.57 & 0.135 $\pm$ 0.003  \\
  \hline
 \multirow{2}{*}{Kosarak}  & sg & 103 & 3 & -1.00 & 2 & 0.012 & 2 & 7.50 $\pm$ 0.55 & 0.035 $\pm$ 0.002  \\ 
   & cbow & 171 & 7 & 0.04 & 10 & 0.001 & 6 & 4.01 $\pm$ 1.79 & 0.021 $\pm$ 0.006 \\ 
   \hline
 \multirow{2}{*}{Twitter follow}  & sg & 50 & 6 & 0.81 & 2 & 0.100 & 21 & 13.01 $\pm$ 0.19 & 0.073 $\pm$ 0.002  \\
  & cbow & 34 & 44 & 0.47 & 8 & 0.074 & 17 & 12.48 $\pm$ 0.52 & 0.070 $\pm$ 0.001  \\
  \hline
 \multirow{2}{*}{Twitter RT}& sg  & 35 & 10 & -0.52 & 8 & 0.100 & 7 & 4.03 $\pm$ 0.04 & 0.020 $\pm$ 0.001 \\
   & cbow  & 183 & 9 & -0.19 & 1 & 0.034 & 1 & 5.51 $\pm$ 0.05 & 0.028 $\pm$ 0.001 \\
 \hline
 \end{tabular}
 \caption{Optimal hyperparameters estimated from a 10\% sample of the dataset with runtime constrained to the runtime on the 10\% sample using the default parameters  $d=100,L=5,\alpha=0.75, n=5, \lambda=0.025$. HR and NDCG are measured by applying the parameters to the full dataset. 95\% confidence intervals reported for HR and NDCG over five runs.  \label{tab:sample}}
\end{center}
\end{table*}

\begin{figure*}
\textbf{Number of Negative Samples}
\vspace{-3mm}
\begin{multicols}{3}
    \includegraphics[width=\linewidth]{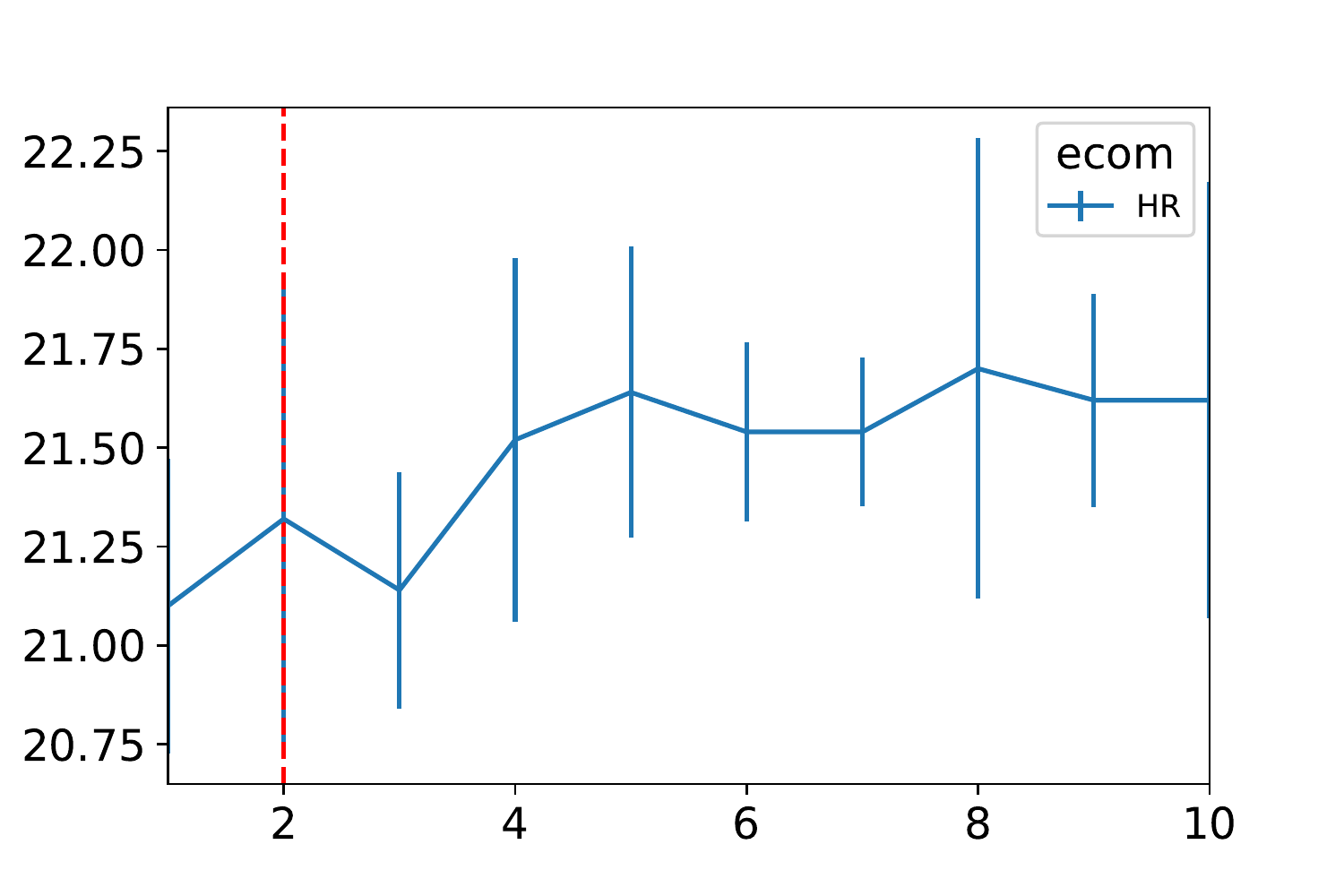} \par
    \includegraphics[width=\linewidth]{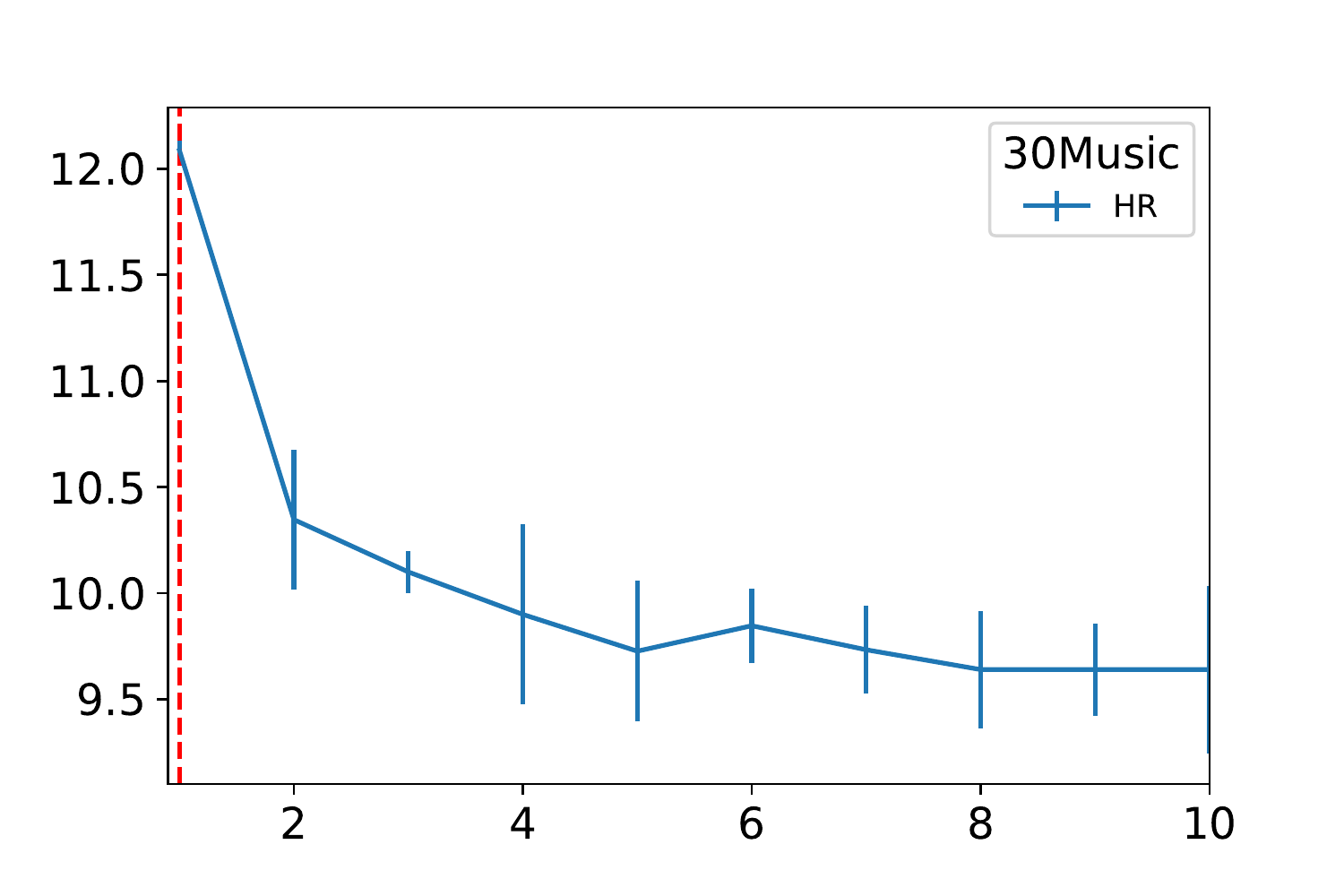} \par
    \includegraphics[width=\linewidth]{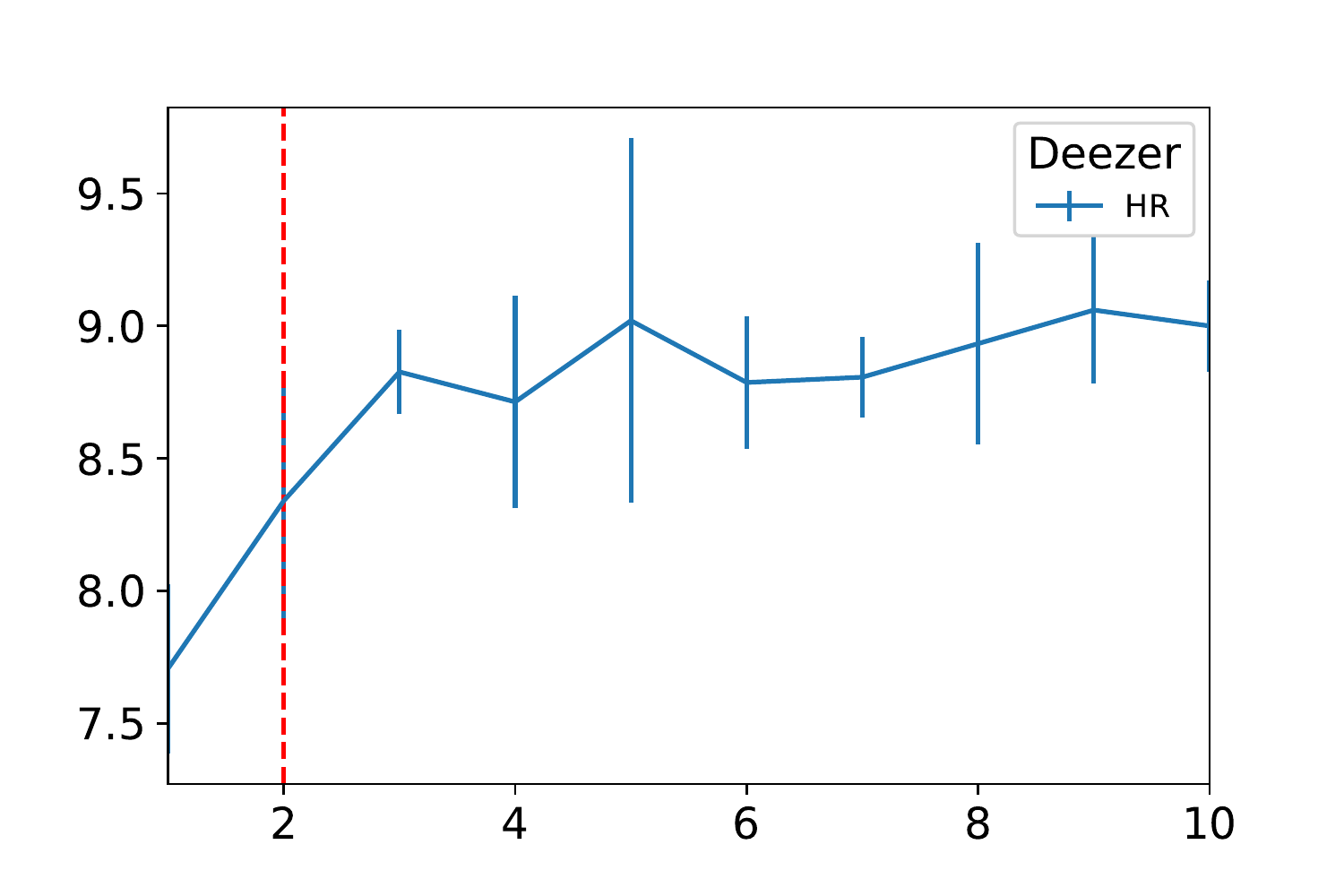} \par
\end{multicols}
\vspace{-8mm}
\begin{multicols}{3}
    \includegraphics[width=\linewidth]{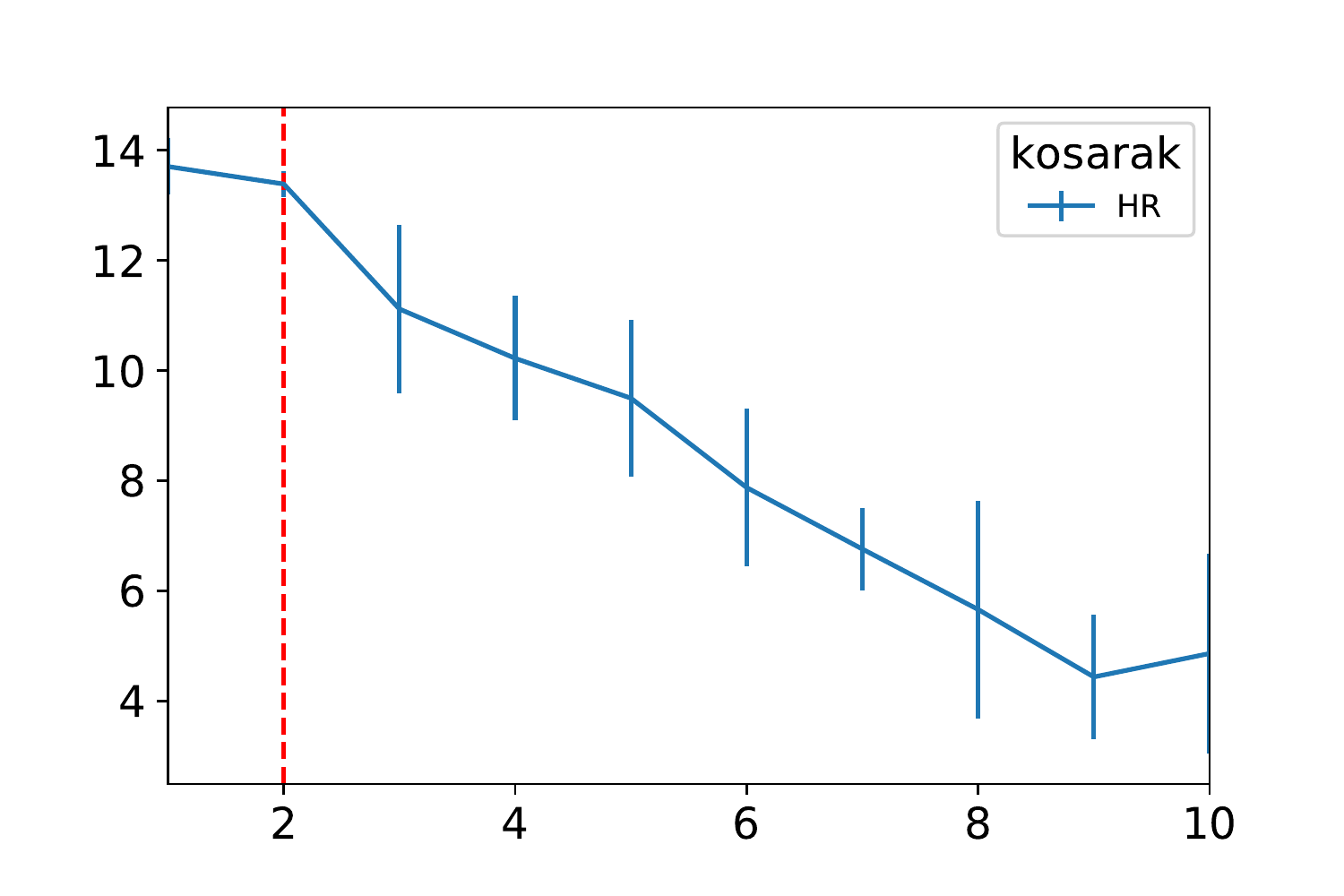} \par
    \includegraphics[width=\linewidth]{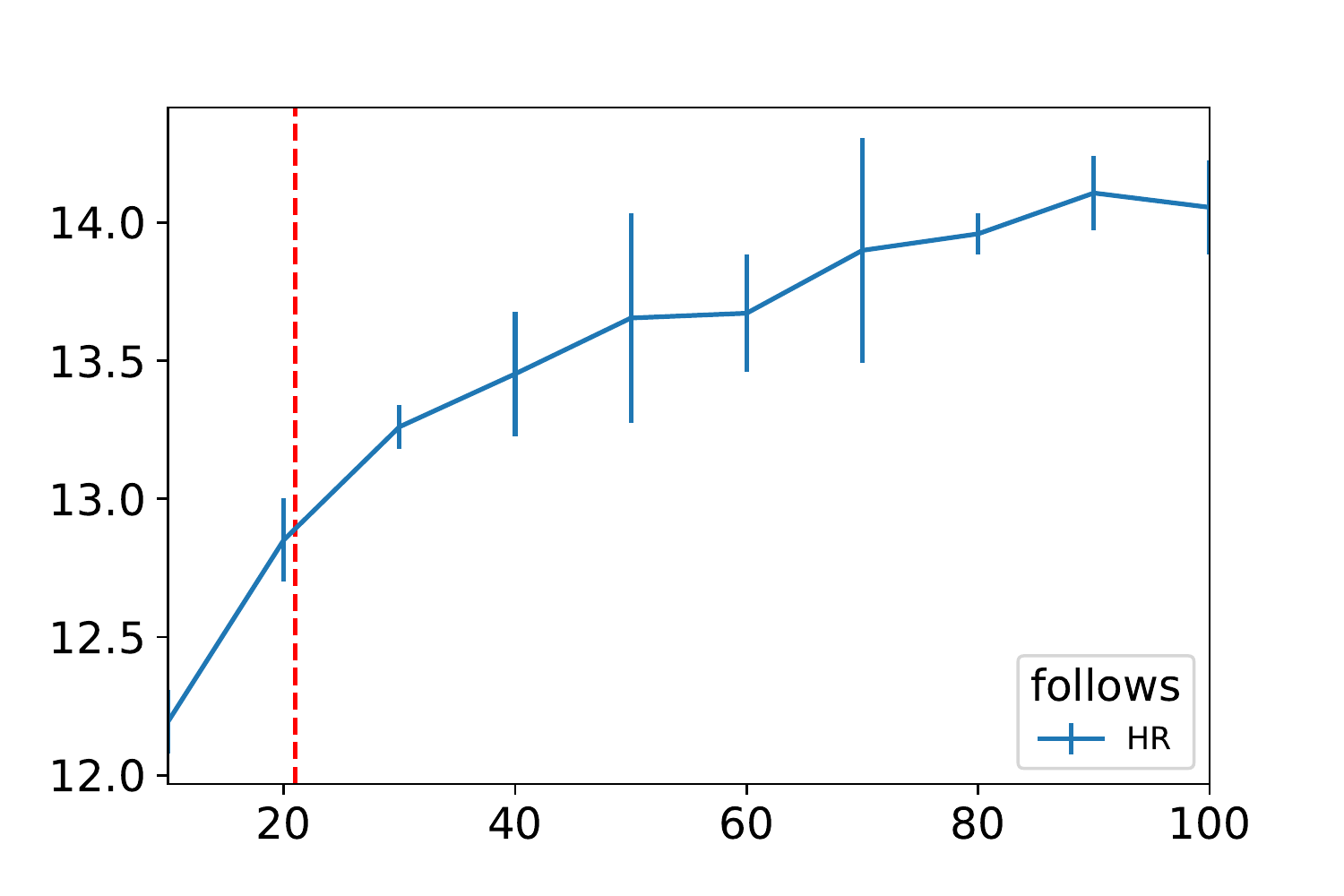} \par
    \includegraphics[width=\linewidth]{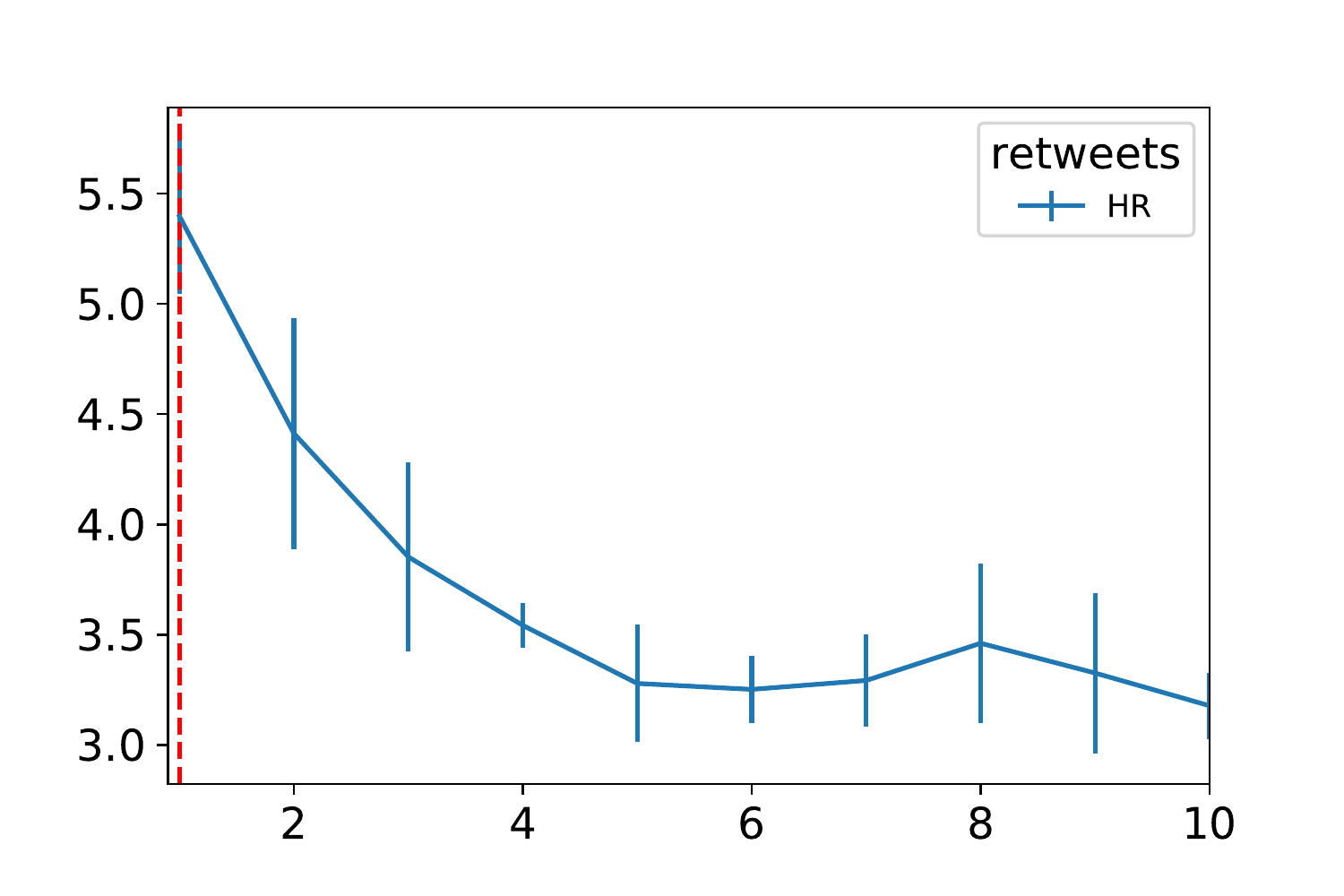} \par
\end{multicols}
\vspace{-3mm}

\textbf{Window Size}
\vspace{-3mm}
\begin{multicols}{3}
    \includegraphics[width=\linewidth]{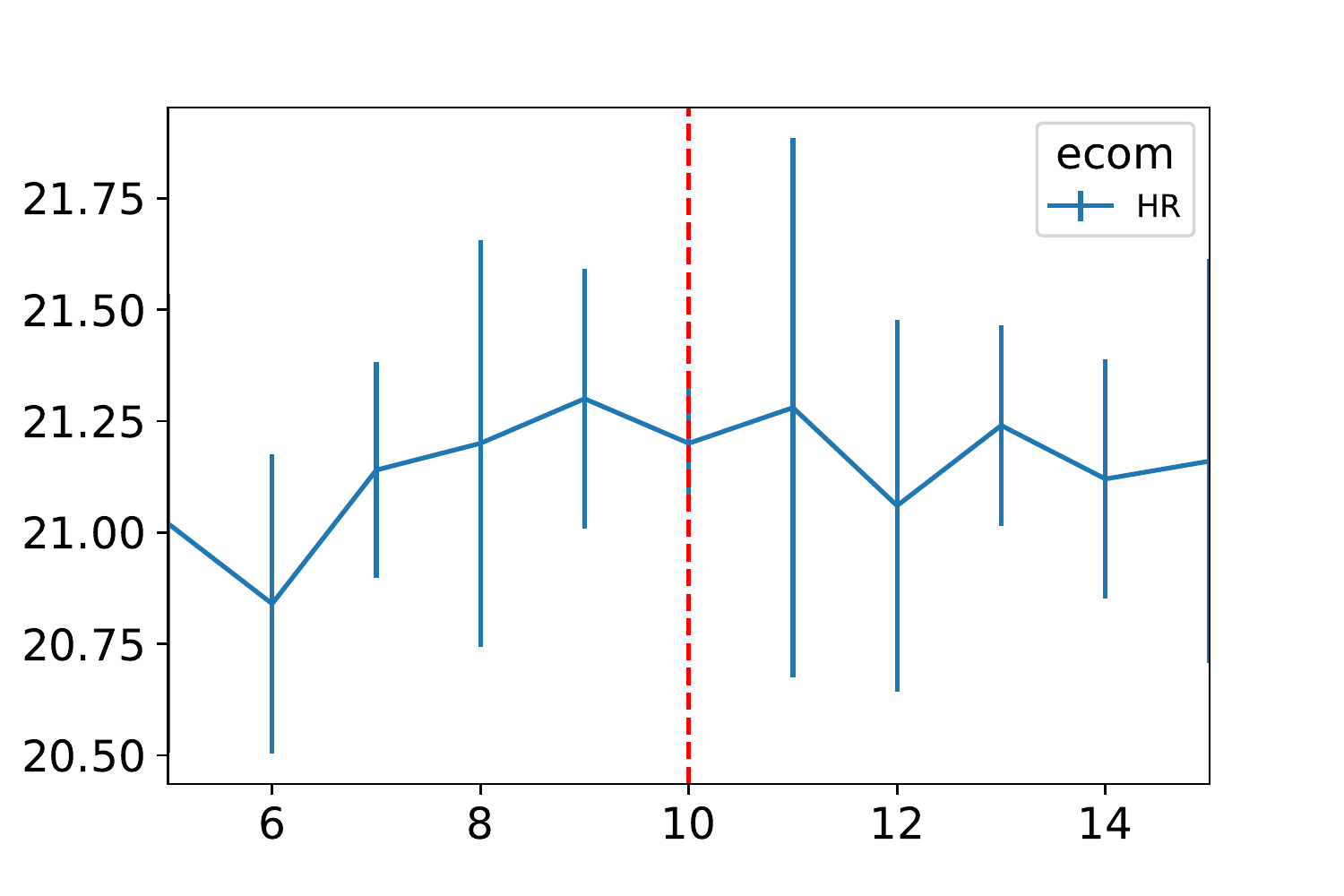} \par
    \includegraphics[width=\linewidth]{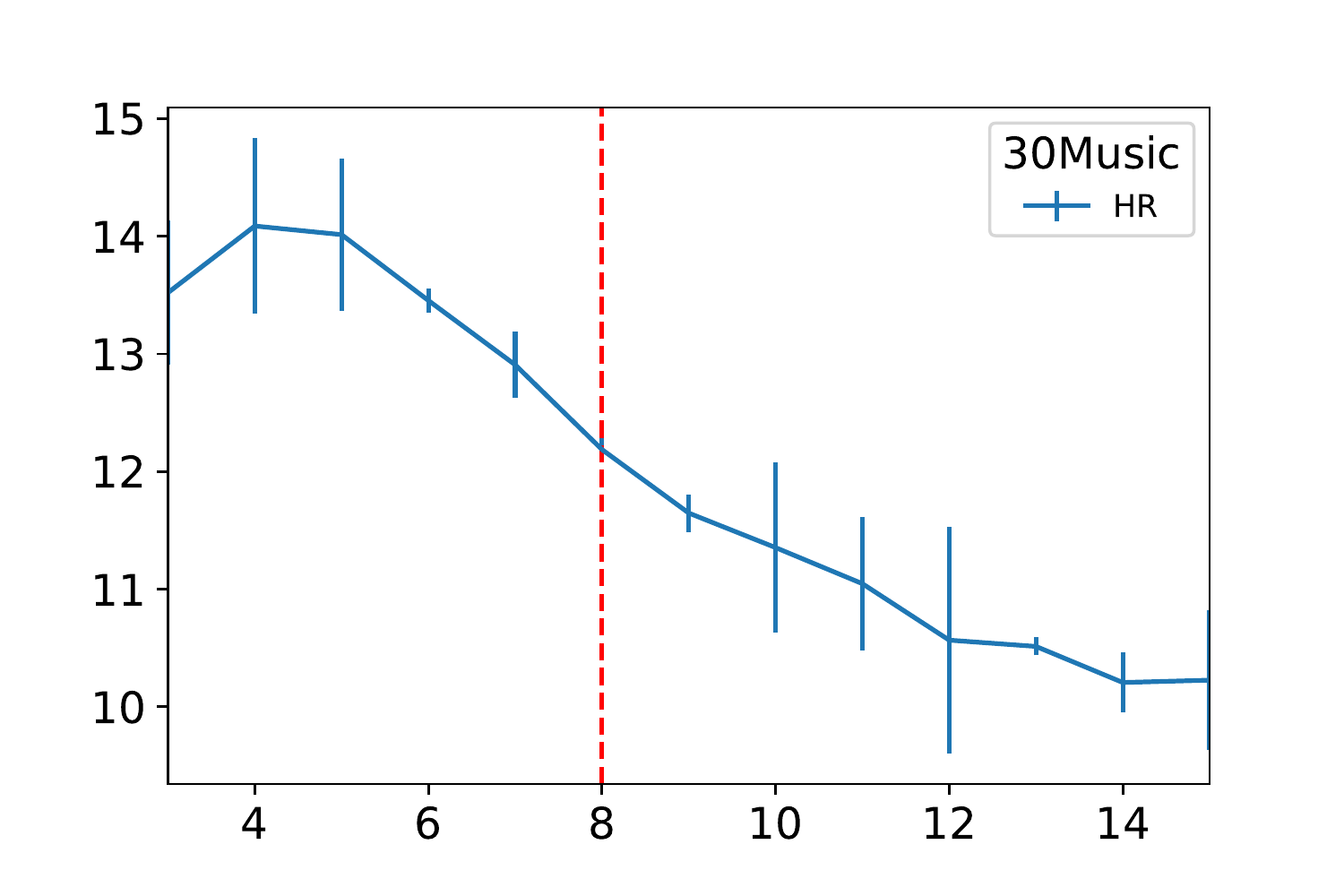} \par
    \includegraphics[width=\linewidth]{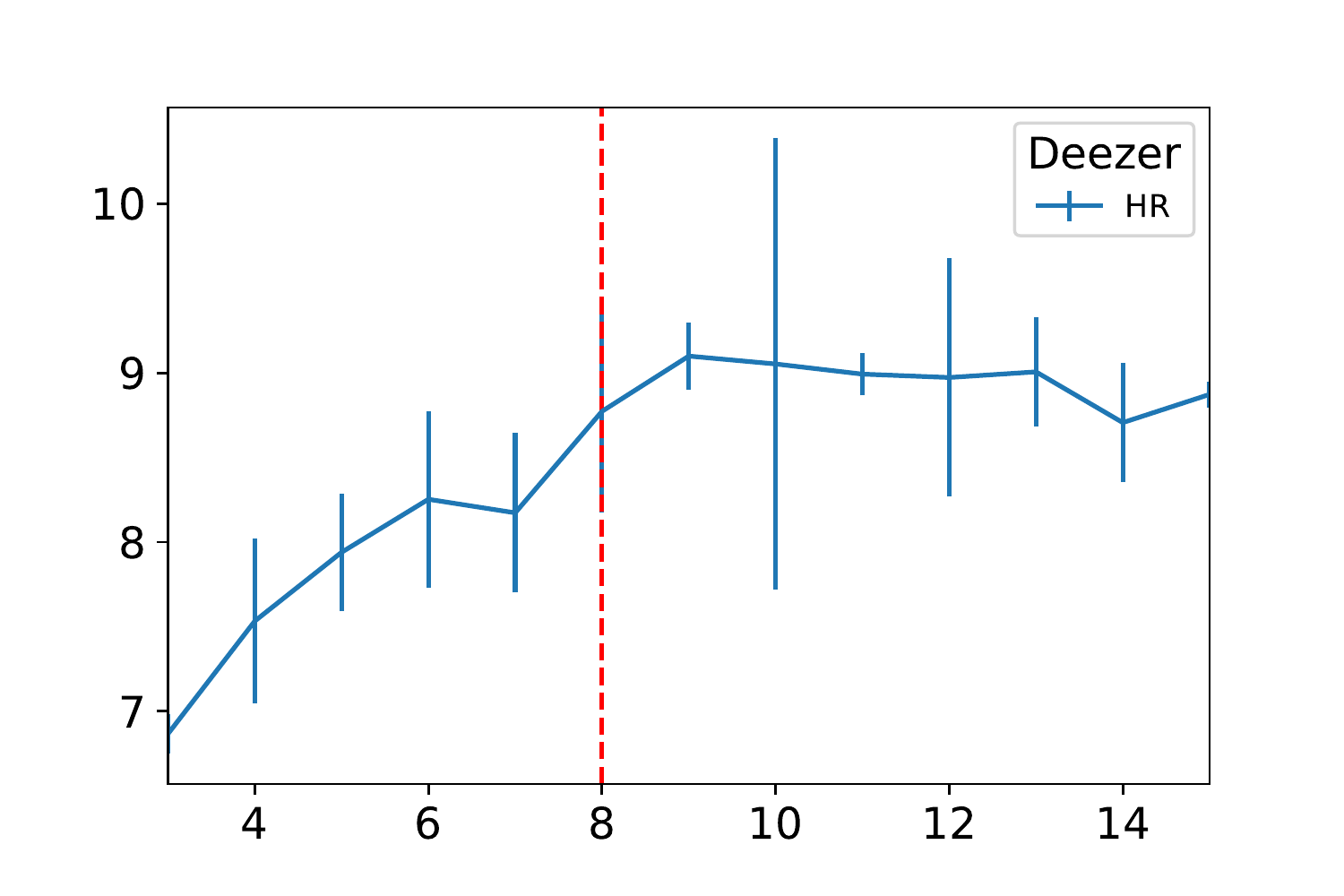} \par
\end{multicols}
\vspace{-8mm}
\begin{multicols}{3}
    \includegraphics[width=\linewidth]{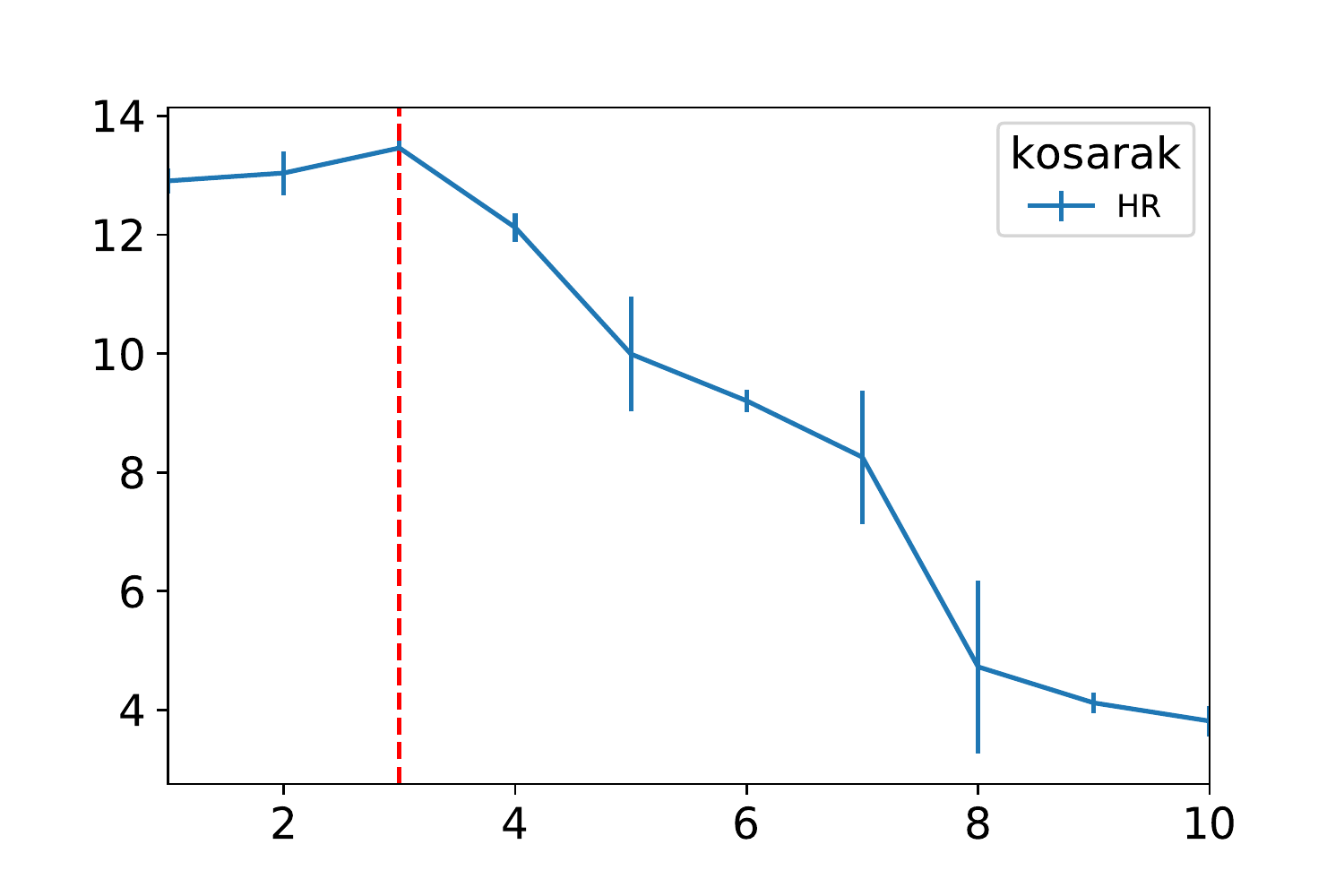} \par
    \includegraphics[width=\linewidth]{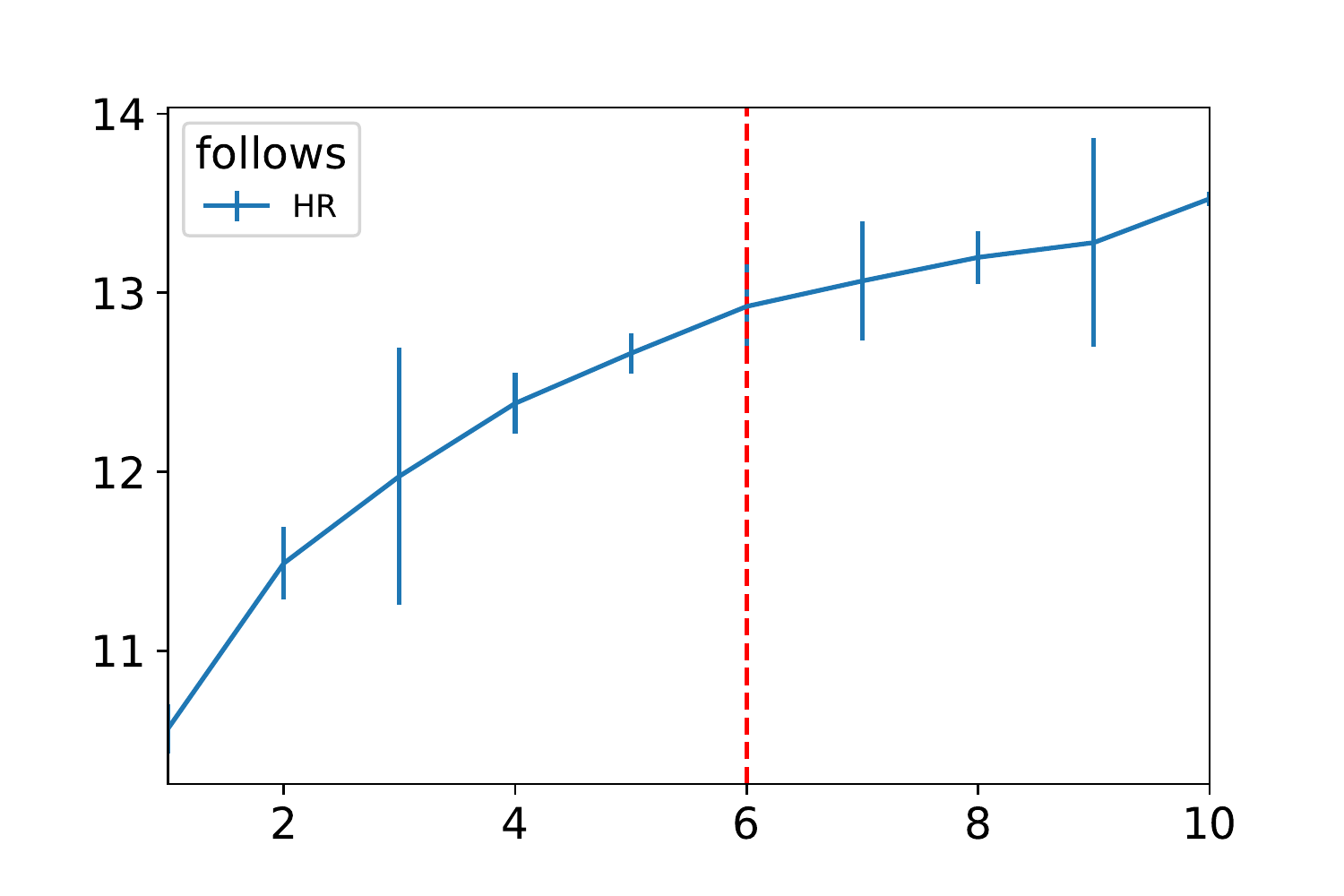} \par
    \includegraphics[width=\linewidth]{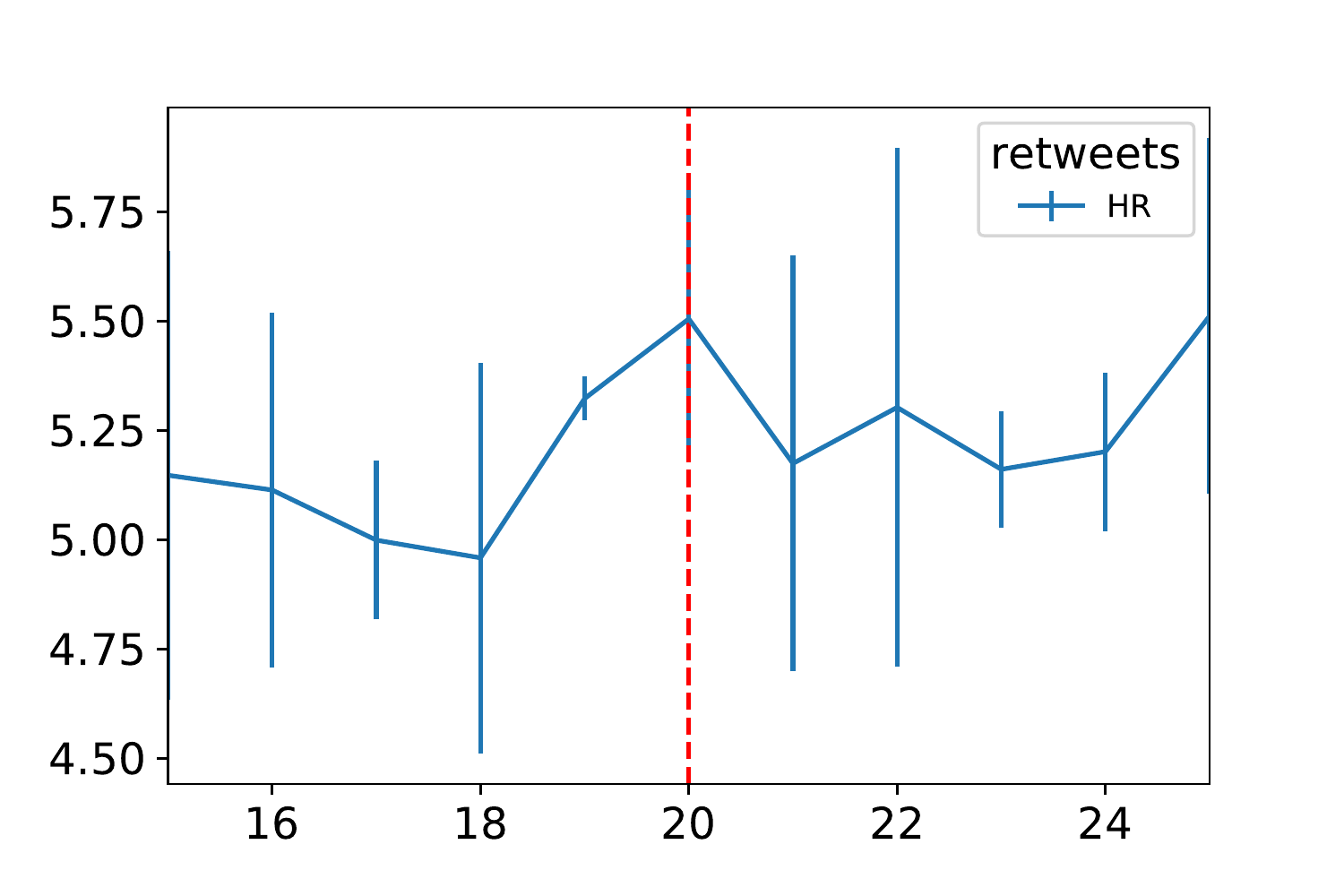} \par
\end{multicols}

\textbf{Embedding Dimension}
\begin{multicols}{3}
    \includegraphics[width=\linewidth]{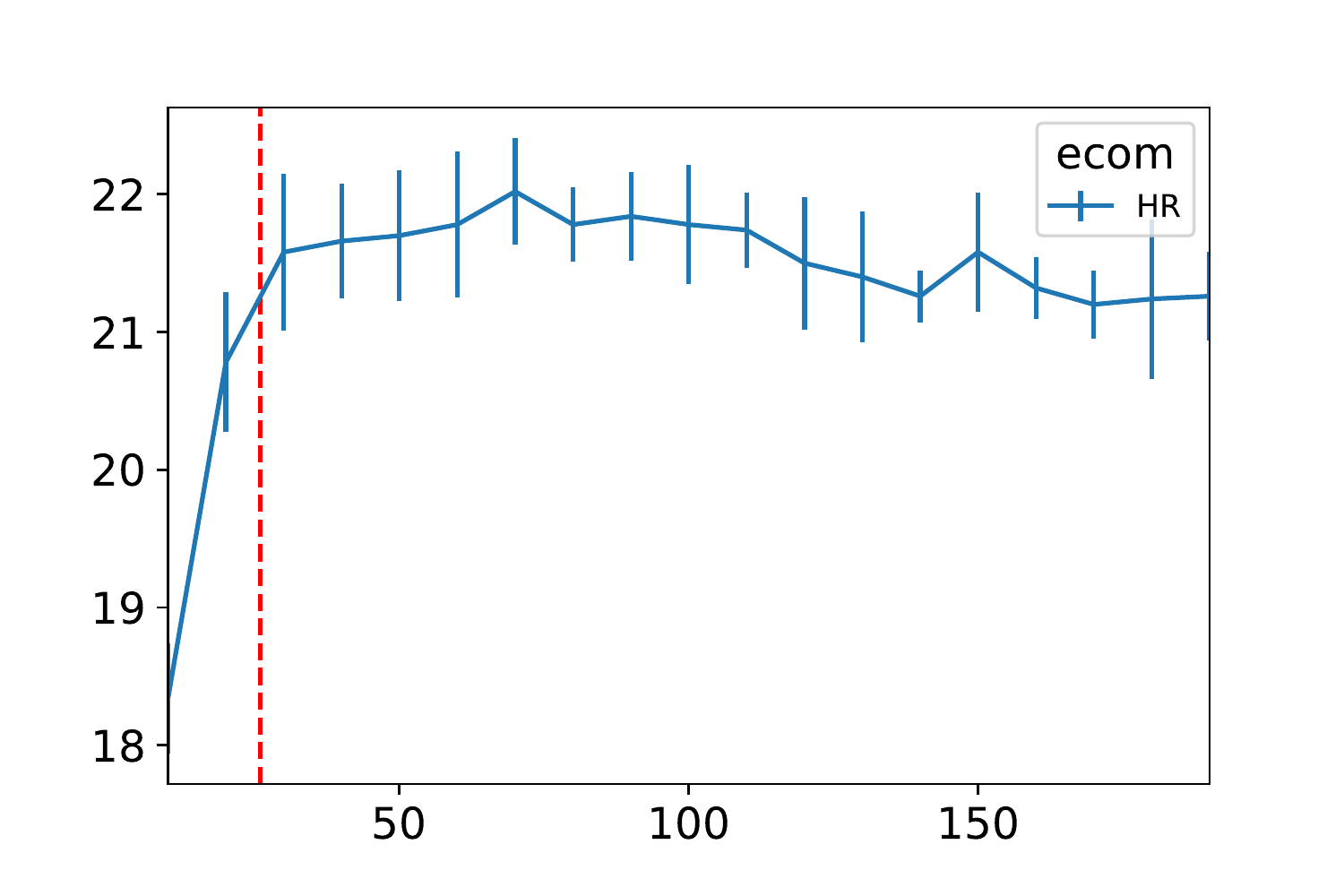}\par
    \includegraphics[width=\linewidth]{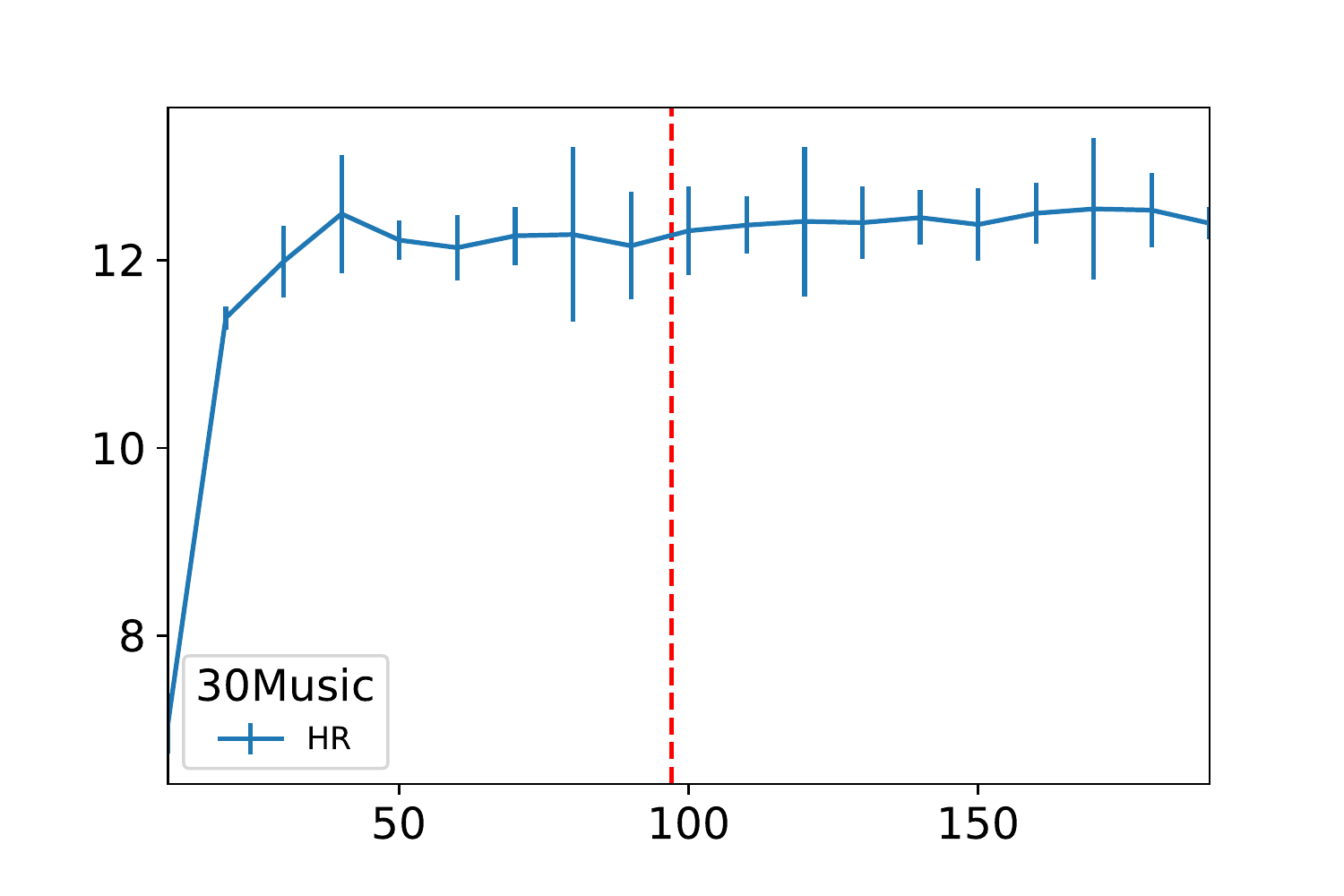}\par
    \includegraphics[width=\linewidth]{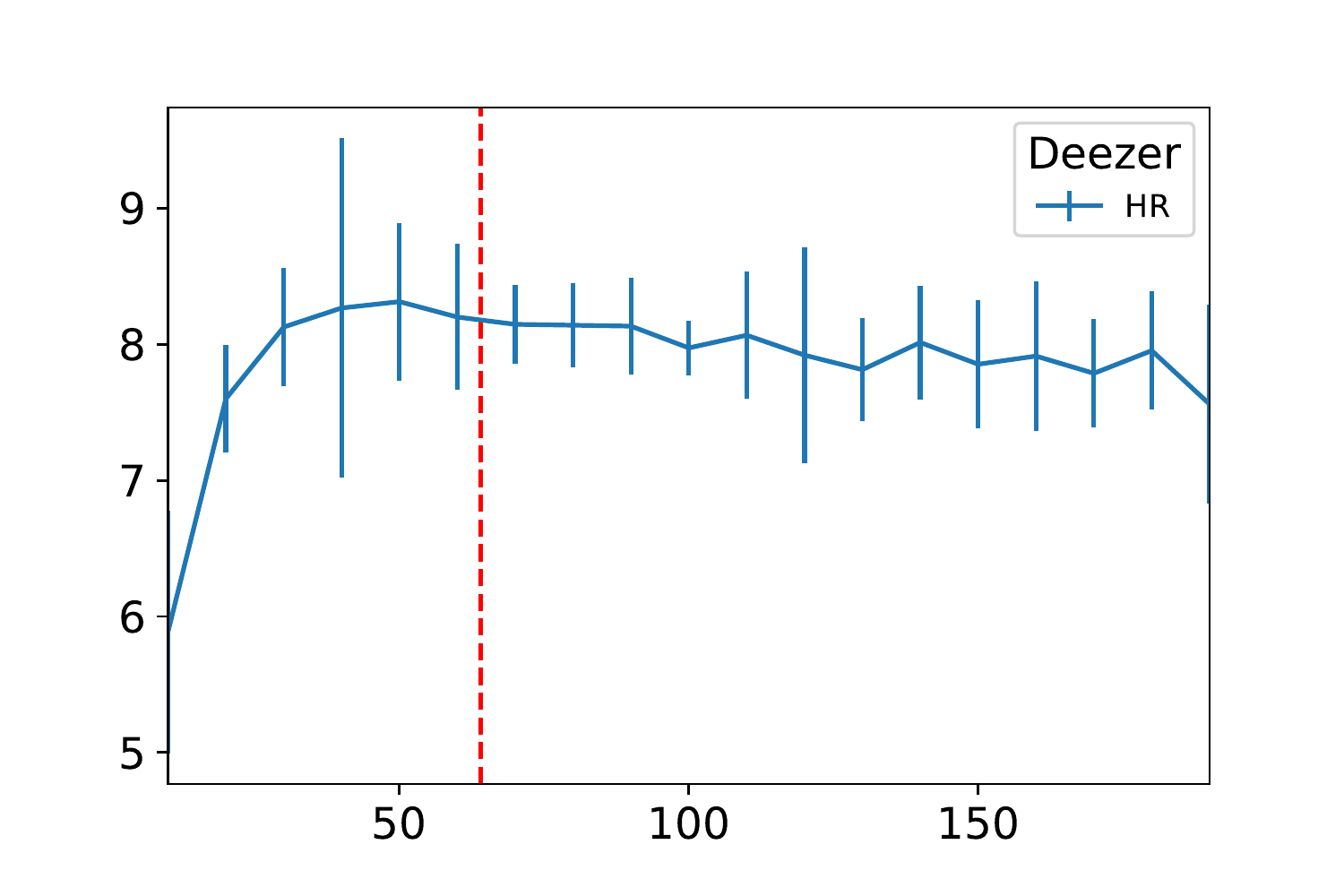}\par
\end{multicols}
\vspace{-8mm}
\begin{multicols}{3}
    \includegraphics[width=\linewidth]{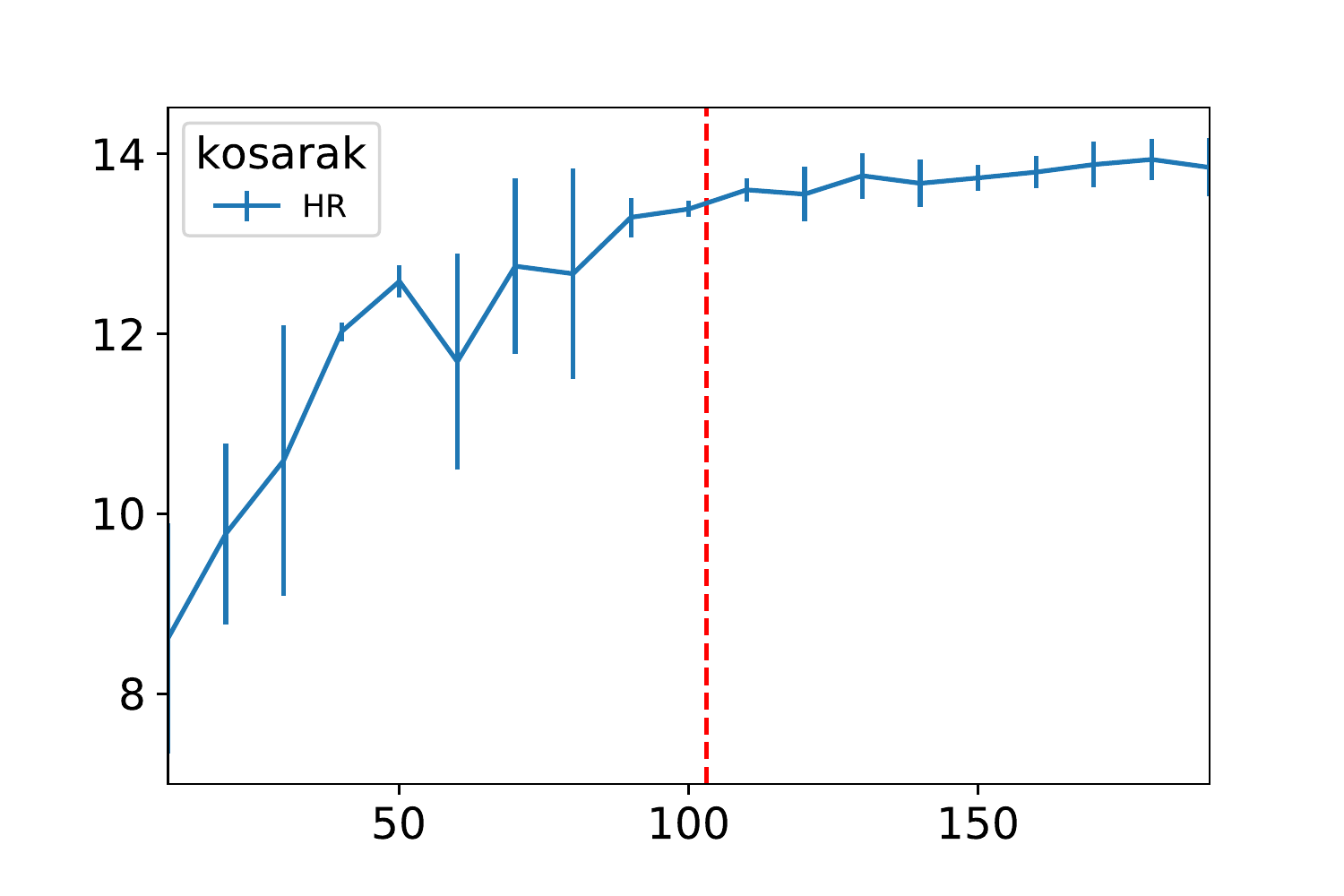}\par
    \includegraphics[width=\linewidth]{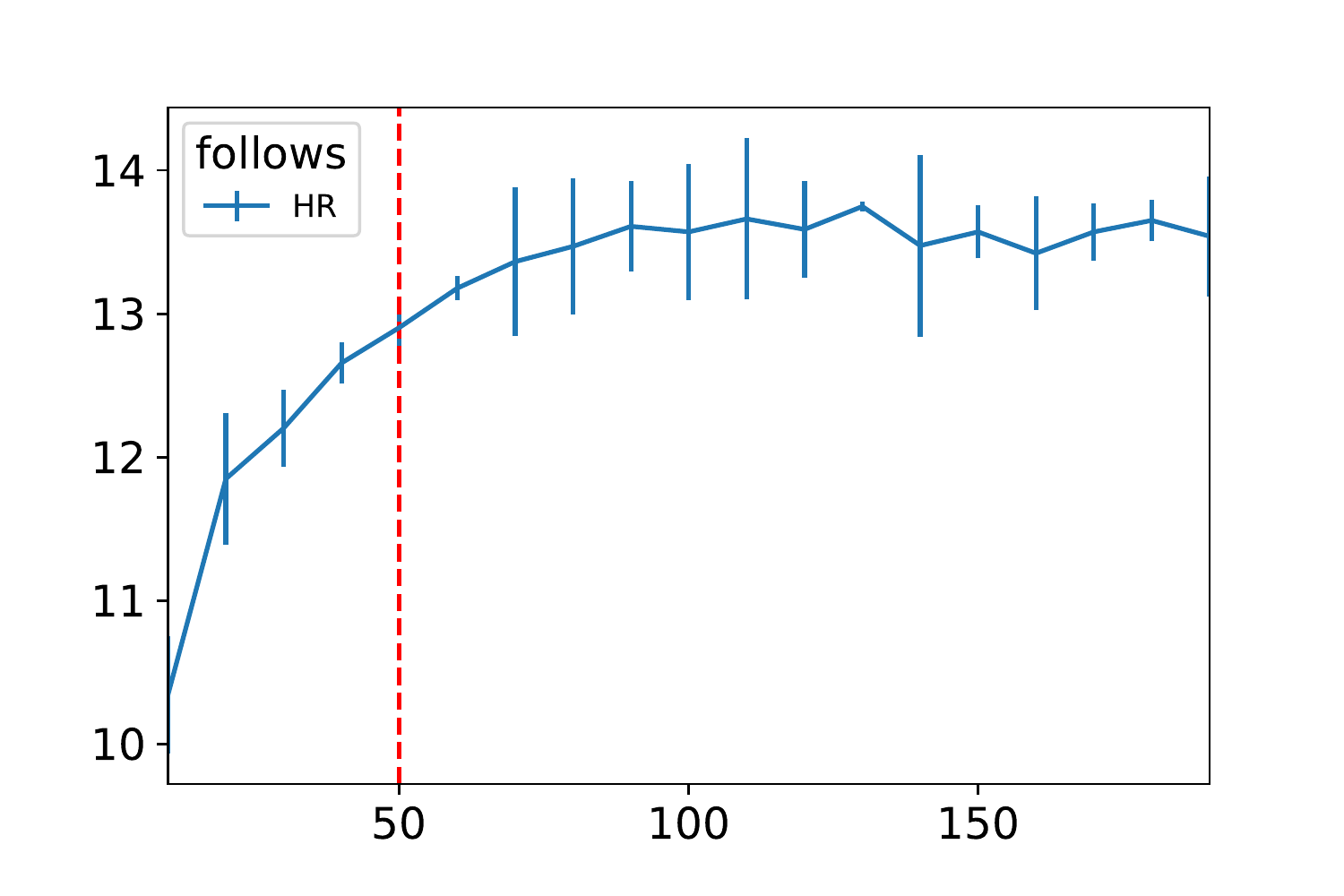}\par
    \includegraphics[width=\linewidth]{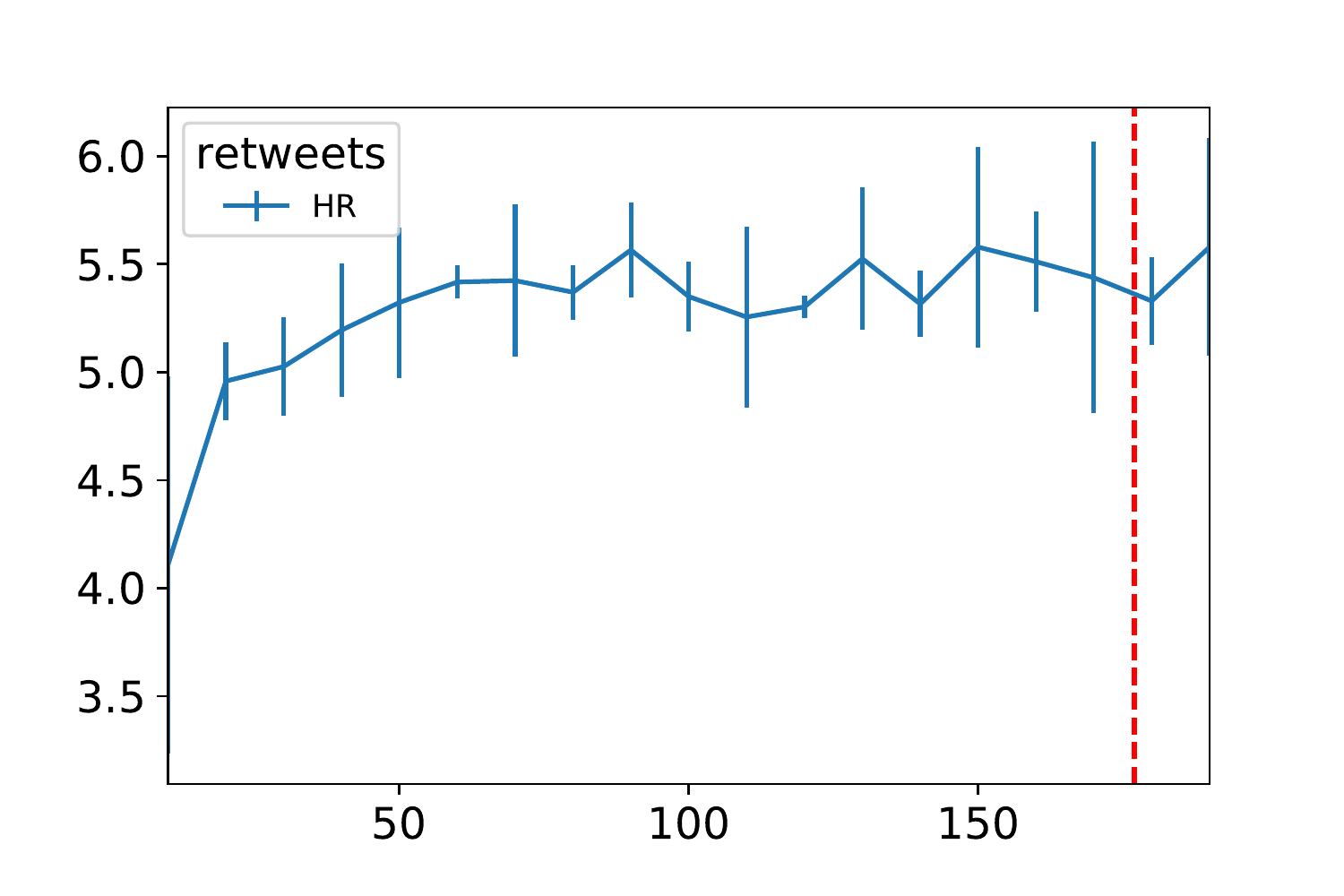}\par
\end{multicols}
\vspace{-7mm}
\caption{HR@10 against negative sampling exponent on full datasets. Red dashed line is the optimal setting using the sample \label{fig:line_plots}}
\end{figure*}

Figure~\ref{fig:line_plots} shows a linear search around the optimal parameters estimated from the sample, but evaluated on the full dataset. The dashed red lines indicate the optimal value found from the sample. The plots show that given a set of hyperparameters discovered using a sample, a small change to a single hyperparameter will not lead to large improvements in performance.  
The optimal negative sampling exponent is consistent between the sample and the full dataset and, unlike the NLP domain, for all datasets, choosing $\alpha=0$ is either optimal or near optimal implying that uniform negative sampling is a good choice for recommender systems. 
We expected to see less negative samples and higher embedding dimension lead to improved performance on the larger dataset compared to the sample. We observe this pattern weakly for the embedding dimension, but not for negative samples. This may be because the constrained optimizer on the sample is already choosing very few negative samples (one or two) for five of the six datasets.

\begin{figure*}[t]
\textbf{Negative sampling exponent}
\vspace{-3mm}
\begin{multicols}{3}
    \includegraphics[width=\linewidth]{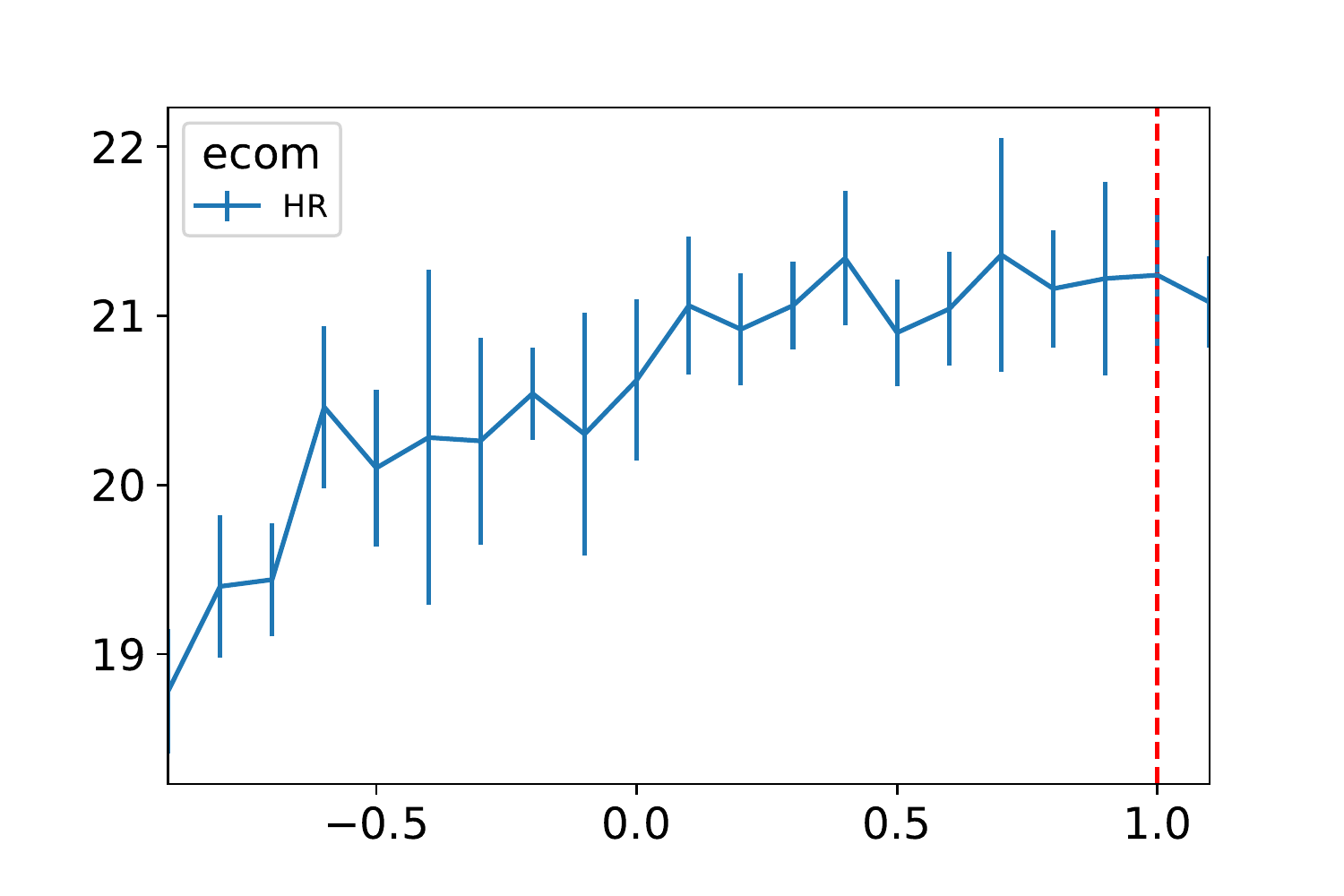}\par
    \includegraphics[width=\linewidth]{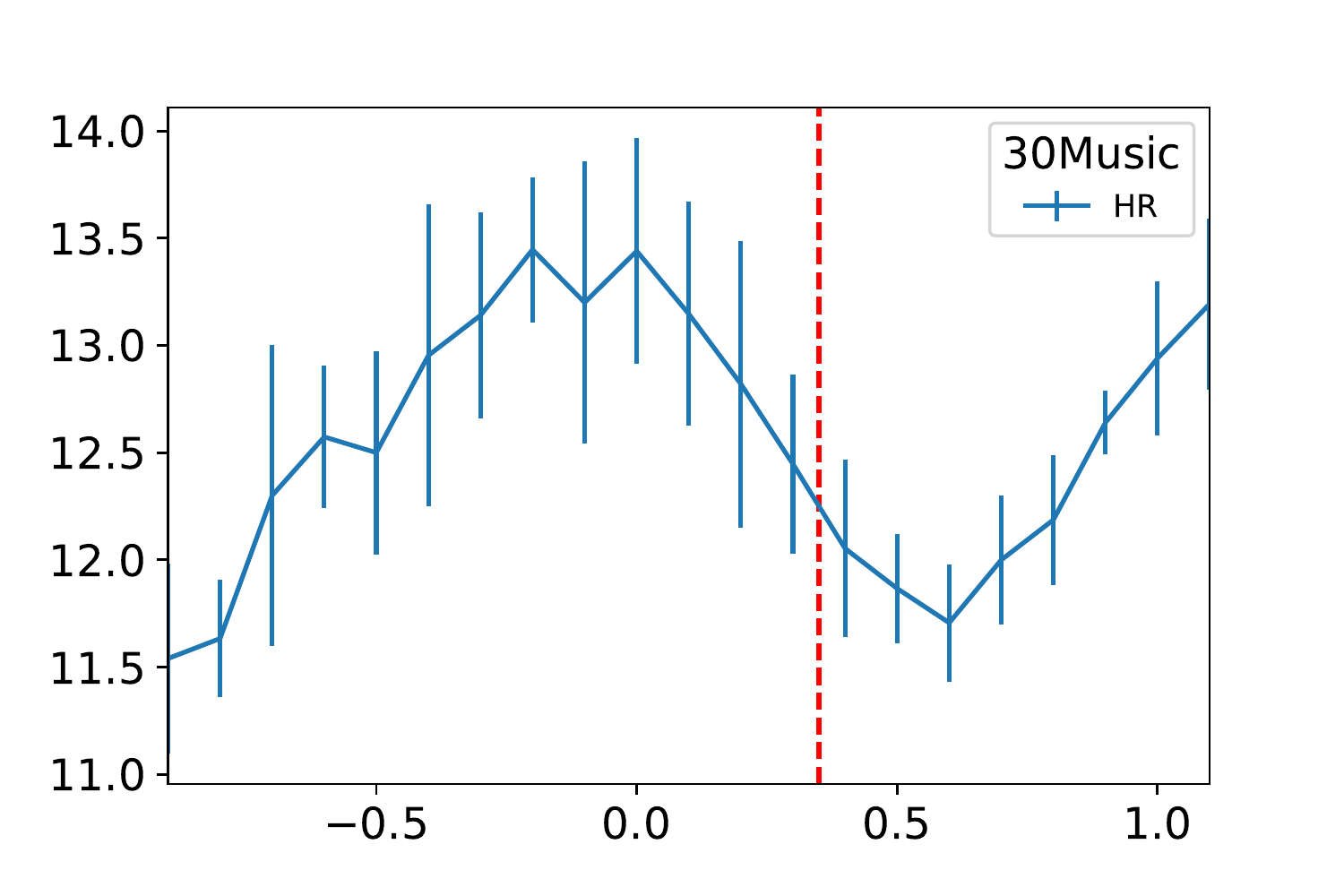}\par
    \includegraphics[width=\linewidth]{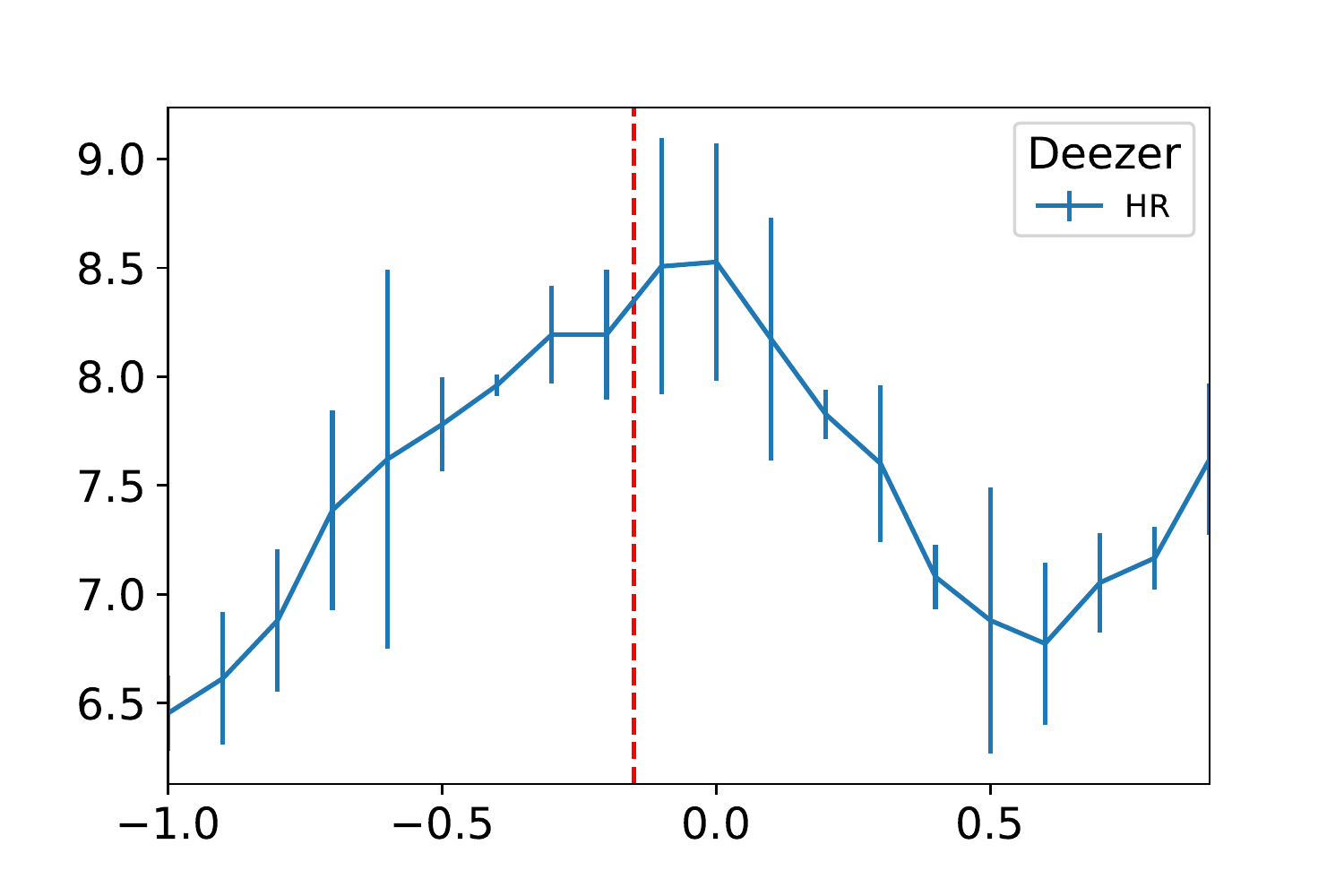}\par
\end{multicols}
\vspace{-8mm}
\begin{multicols}{3}
    \includegraphics[width=\linewidth]{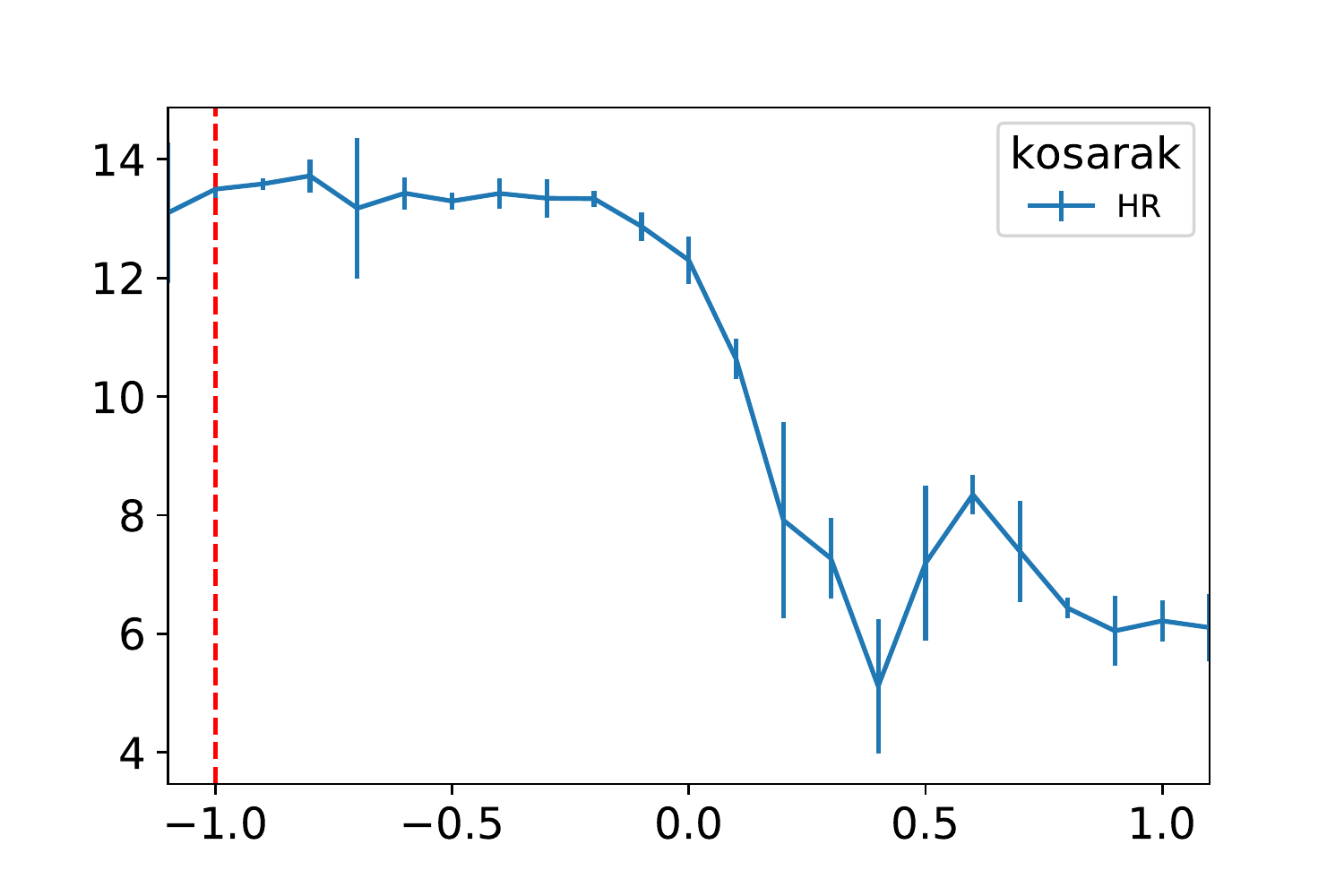}\par
    \includegraphics[width=\linewidth]{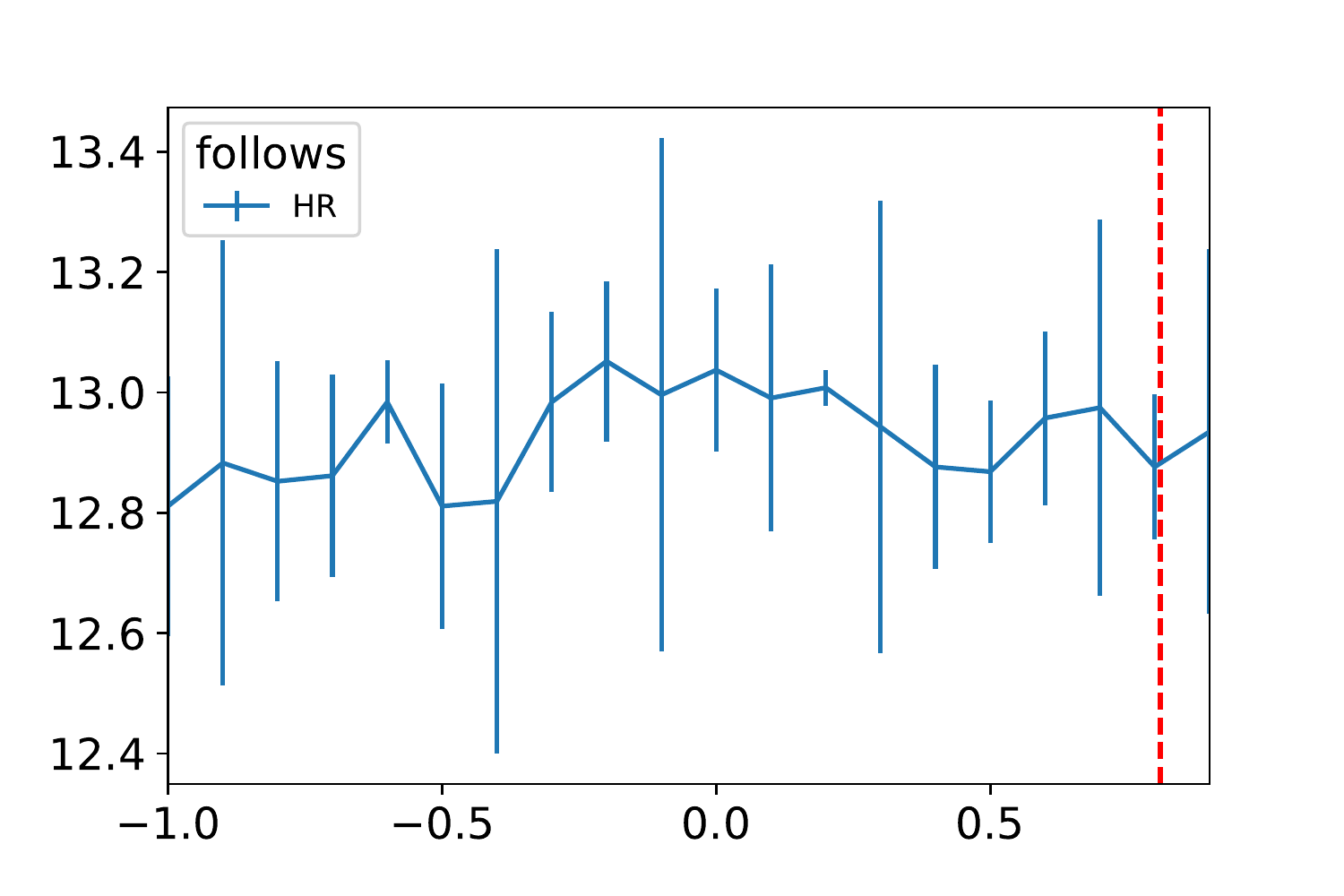}\par
    \includegraphics[width=\linewidth]{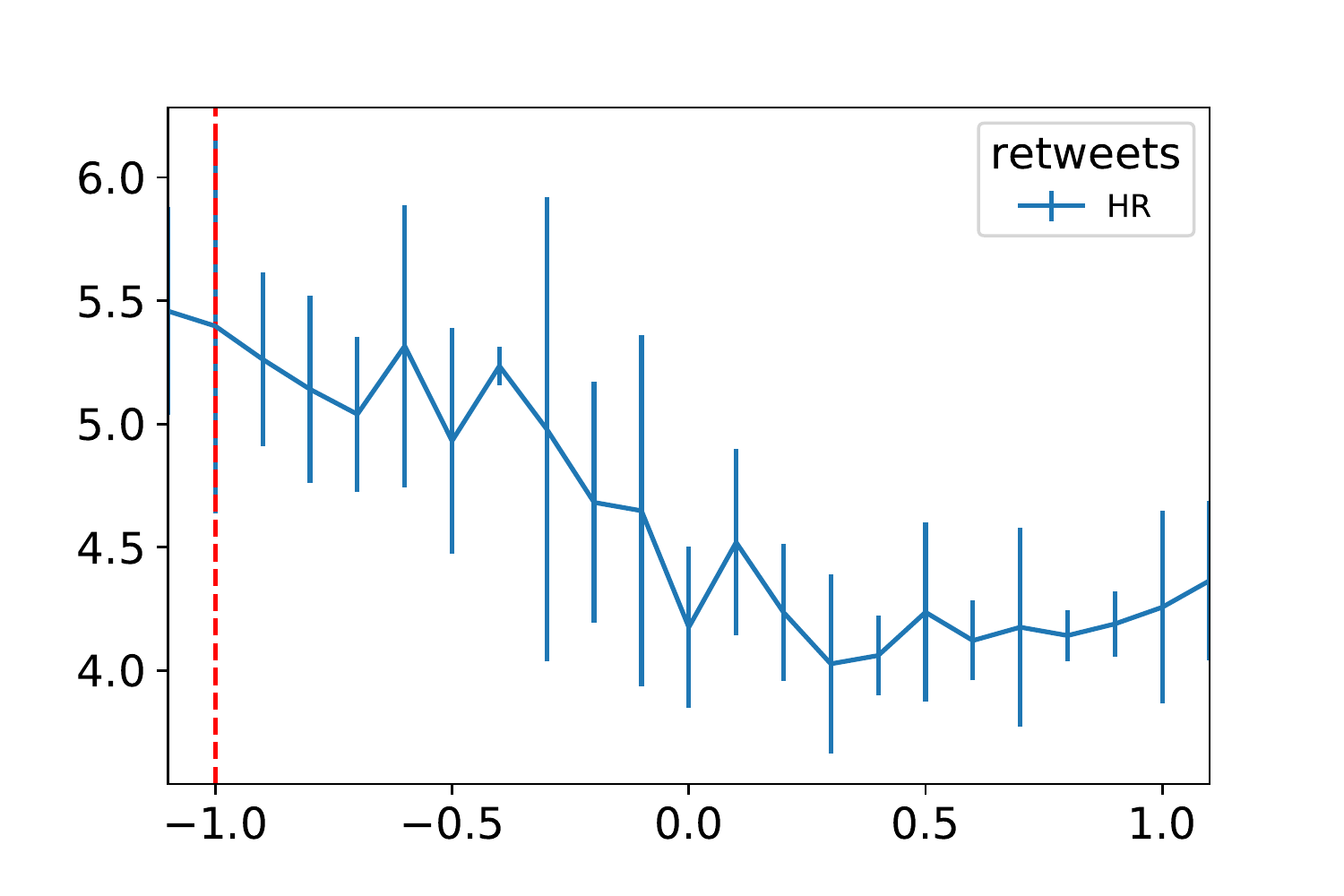}\par
\end{multicols}
\vspace{-3mm}
\textbf{Initial Learning Rate}
\vspace{-3mm}
\begin{multicols}{3}
    \includegraphics[width=\linewidth]{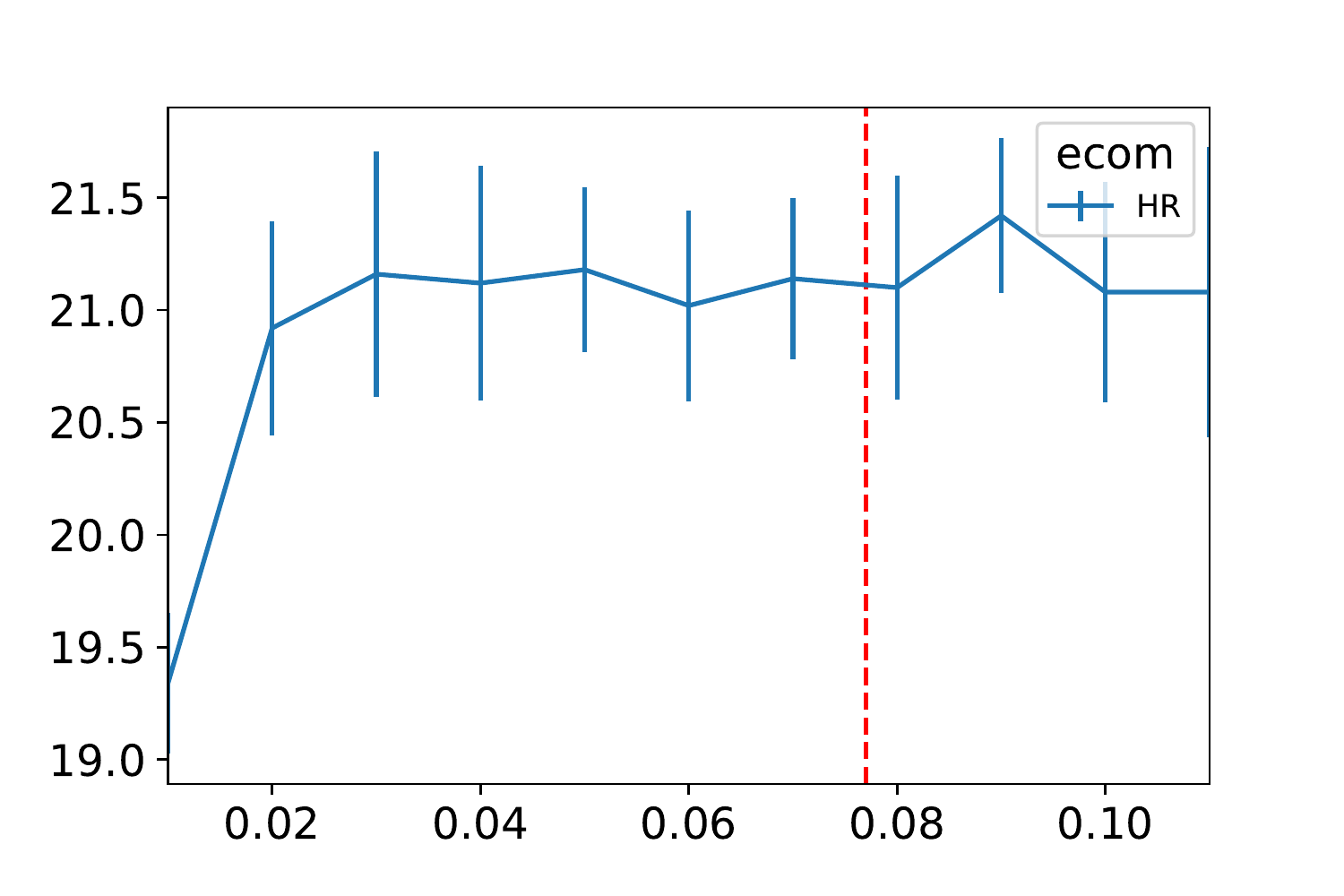} \par
    \includegraphics[width=\linewidth]{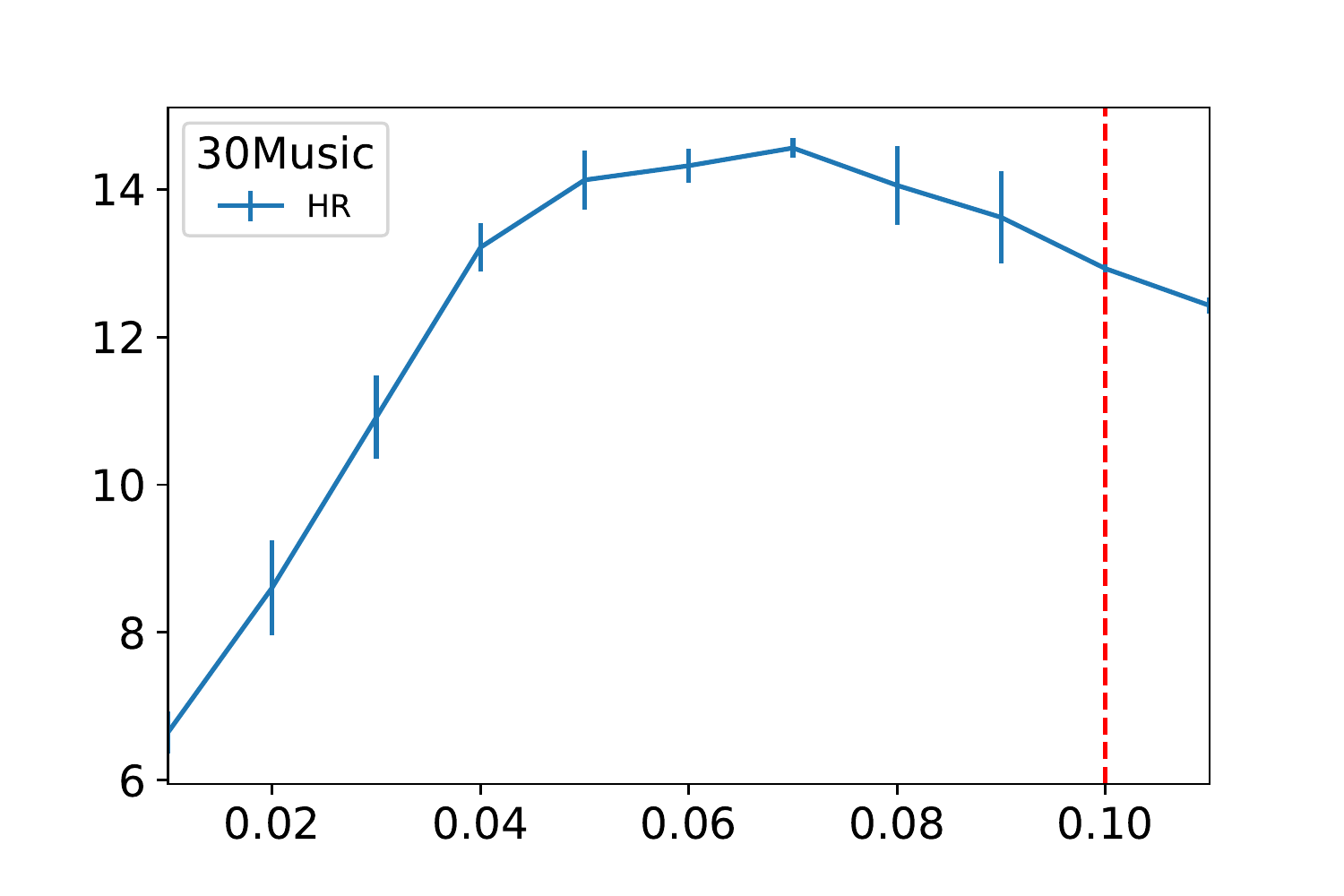} \par
    \includegraphics[width=\linewidth]{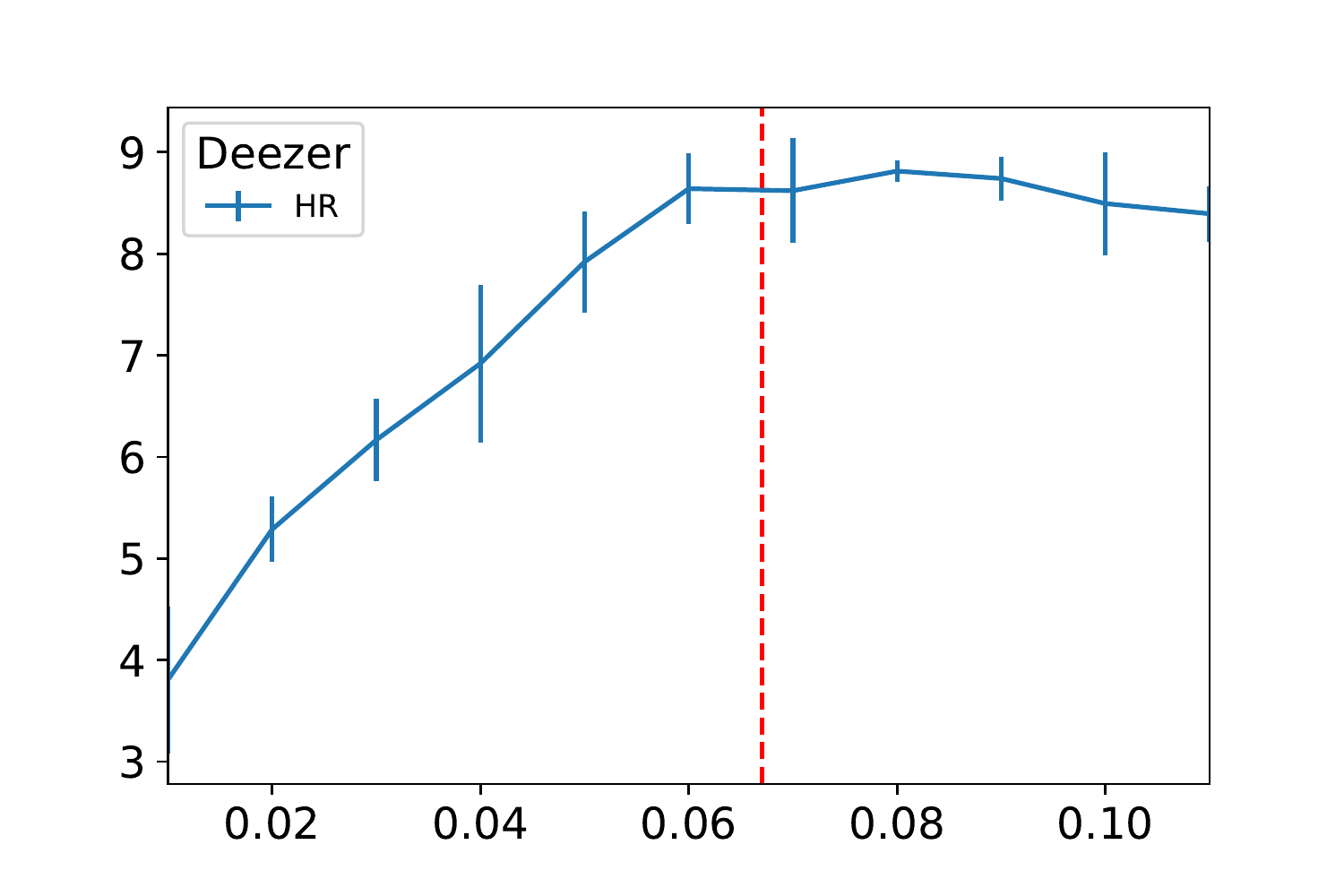} \par
\end{multicols}
\vspace{-8mm}
\begin{multicols}{3}
    \includegraphics[width=\linewidth]{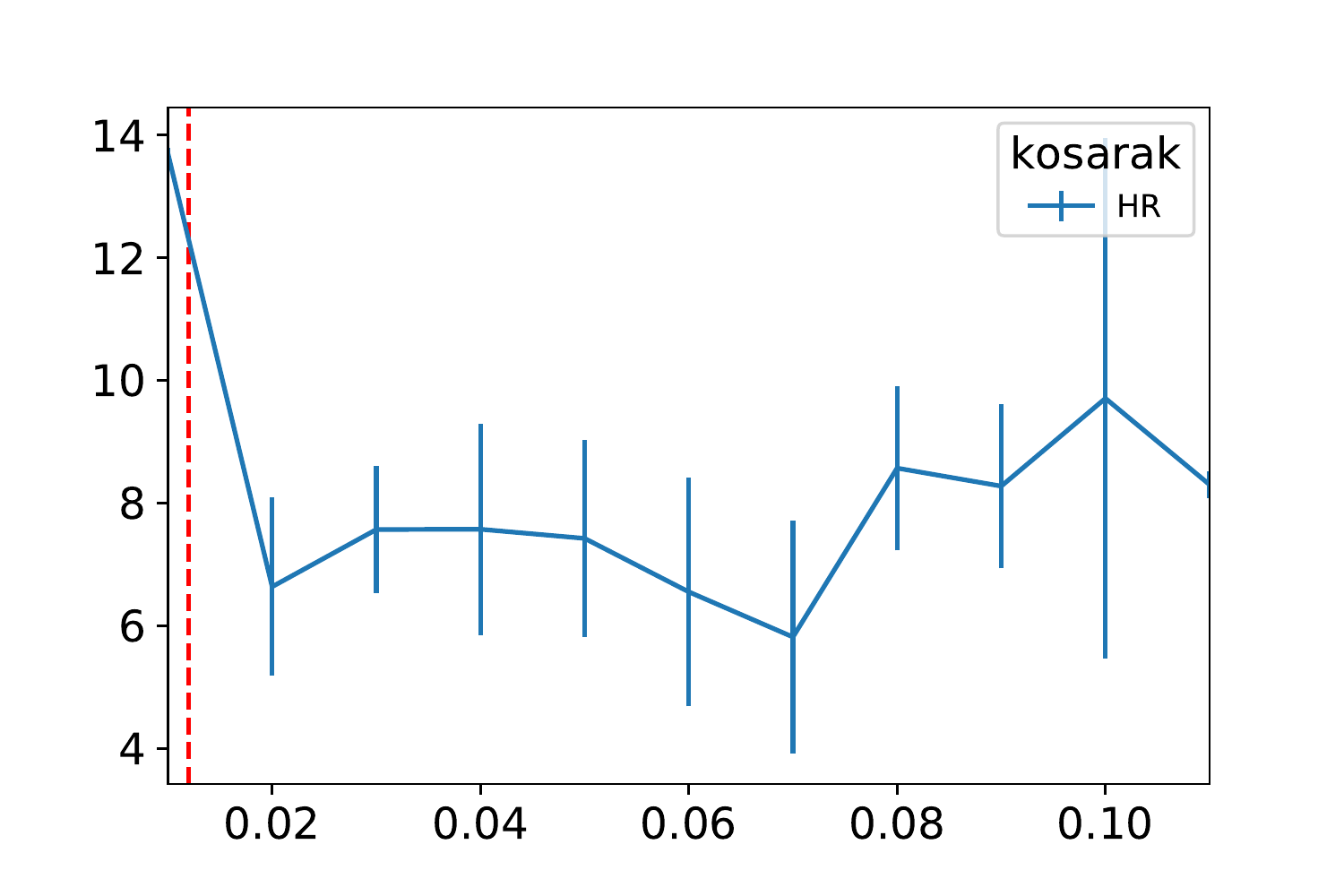} \par
    \includegraphics[width=\linewidth]{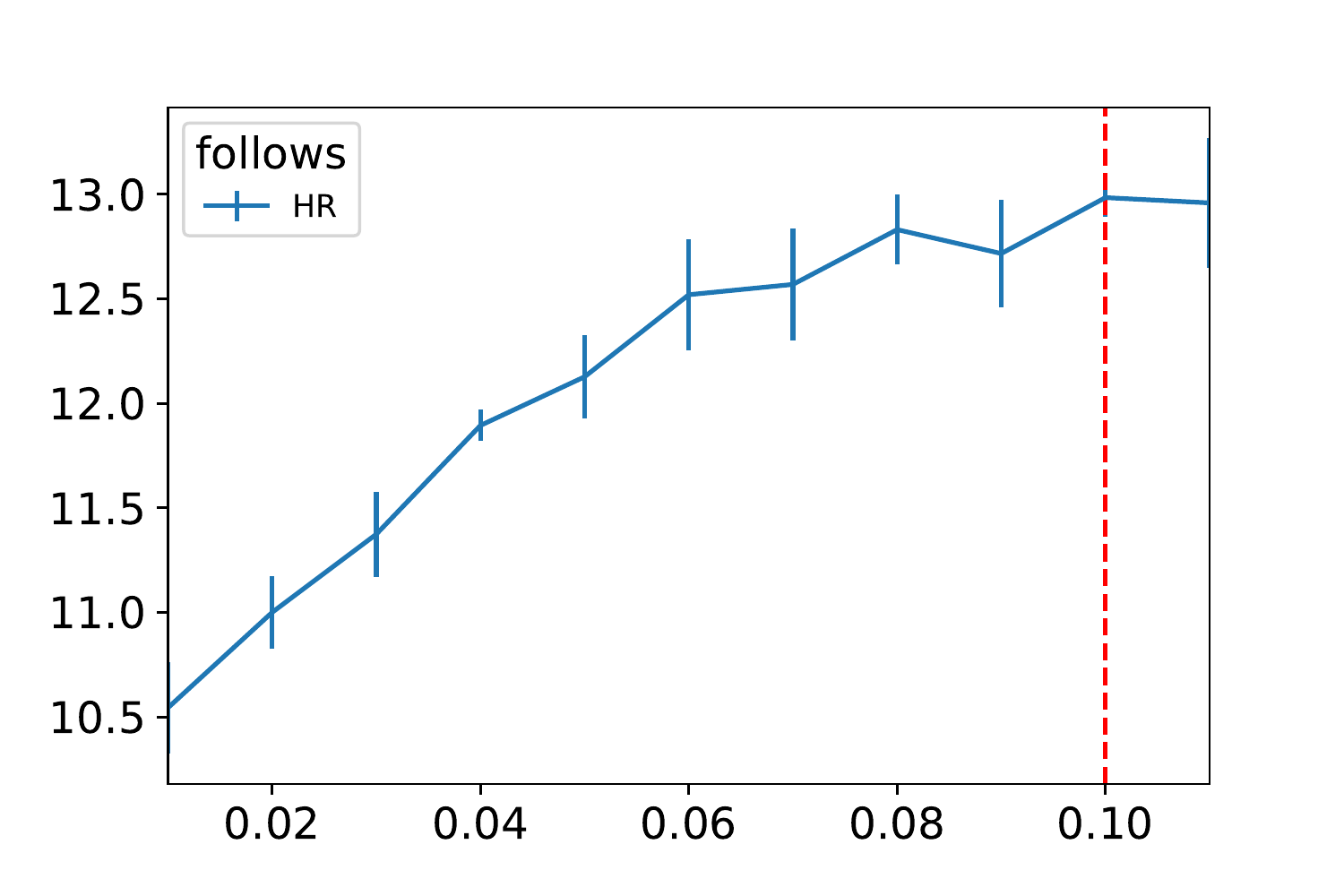} \par
    \includegraphics[width=\linewidth]{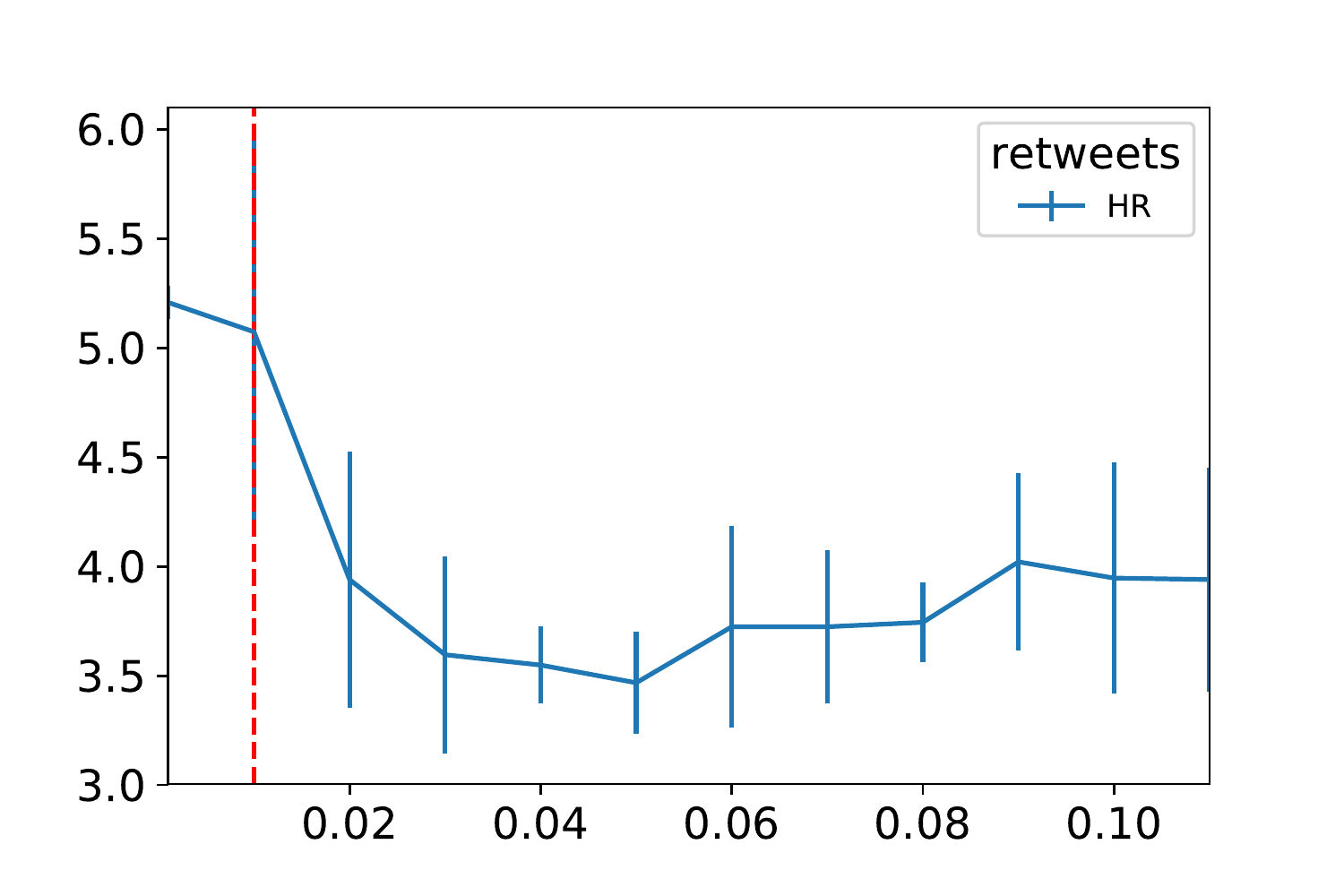} \par
\end{multicols}
\vspace{-3mm}
\end{figure*}

\subsection{Online experiment}
\label{sec:online_experiment}
We evaluate the gain from tuning word2vec for the large scale Who To Follow recommendation task at Twitter~\cite{gupta2013wtf}. The task involves generating top-k similar users as recommendation candidates. 
We trained Word2vec models to learn user embeddings from the sequence of user follows using the default and optimized hyperparameters. Hyperparameters were optimized using a \nicefrac{1}{2000} sample of the full dataset. Furthermore, we generated top-k similar candidates using the approximate nearest neighbor search with the cosine similarity metric. In the online experiment, the hyper-parameter tuned Word2vec model yielded a 14.8\% higher follow rate compared to the default parameters. To our knowledge, this is the largest follow rate gain we have seen by optimizing hyperparameters in our candidate generation algorithm. This result reinforces the importance of Word2vec hyperparameter tuning for the recommendation task.

\section{Conclusion}

Firstly we established that large improvements in recommendation performance can be achieved by optimizing the Word2vec hyperparameters. This result is difficult to translate into large-scale systems, where the size of datasets necessitates trade-offs between runtime and performance. Using constrained Bayesian optimization we showed that recommendations performance can be significantly improved without increasing algorithmic runtimes. Finally, as it is often difficult to run hyperparameter optimization directly on large scale datasets, we showed that to a large extent of the performance boost is achieved by optimizing on a sample, which we demonstrate offline and online testing.

\section{Acknowledgments}

We are grateful to Apoorv Sharma, Christine Chen, Jerry Jiang, Max Hansmire, Sumit Binnani and Yao Wu for useful discussions while developing this work. We would also like to thank Gerard Casas Saez for his role in developing Twitter's hyperparameter optimization tooling (Tuner).






\FloatBarrier
\newcommand{\showDOI}[1]{\unskip} 
\newcommand{\showURL}[1]{\unskip} 
\bibliographystyle{ACM-Reference-Format}

\end{document}